\documentclass[10pt,journal]{IEEEtran}

\usepackage{amsmath,amssymb,mathtools,bm}
\usepackage{amsthm}
\usepackage{graphicx}
\usepackage{booktabs}
\usepackage{multirow}
\usepackage{cite}
\usepackage{array}
\usepackage{url}
\usepackage{xcolor}

\usepackage{enumitem}
\usepackage{balance}
\usepackage{microtype}
\usepackage[caption=false,font=footnotesize]{subfig}

\newtheorem{theorem}{Theorem}

\newtheorem{lemma}{Lemma}
\newtheorem{corollary}{Corollary}
\newtheorem{remark}{Remark}

\newcommand{\CN}{\mathcal{CN}}
\newcommand{\E}{\mathbb{E}}

\newcommand{\diag}{\operatorname{diag}}

\newcommand{\cH}{\mathsf{H}}
\newcommand{\cT}{\mathsf{T}}
\newcommand{\eps}{\varepsilon}
\newcommand{\I}{\mathbf I}
\newcommand{\zero}{\mathbf 0}
\newcommand{\mcQ}{\mathcal Q}
\newcommand{\mcD}{\mathcal D}
\newcommand{\mcG}{\mathcal G}
\newcommand{\mcY}{\mathcal Y}
\newcommand{\mcI}{\mathcal I}

\newcommand{\safeincludegraphics}[2][\columnwidth]{%
  \IfFileExists{#2}{\includegraphics[width=#1]{#2}}{%
    \fbox{\parbox[c][3.2cm][c]{0.92\columnwidth}{\centering\footnotesize
    Figure placeholder: \\ \texttt{#2}}}%
  }%
}

\title{Coordinated and Uncoordinated Channel Tracking for High-Rate IoT in Distributed Massive MIMO With Low-Resolution ADCs}

\author{Hediyeh~Soltanizadeh and~Ardavan~Rahimian,~\IEEEmembership{Senior Member,~IEEE}%
\thanks{\textit{(Corresponding author: Ardavan Rahimian.)}}%
\thanks{H.~Soltanizadeh is with the School of Electrical Engineering, Iran University of Science and Technology (IUST), Tehran, Iran
(e-mail: he.soltanizadeh@gmail.com).}%
\thanks{A.~Rahimian is with the School of Engineering, Ulster University, Belfast BT15 1AP, U.K.
(e-mail: a.rahimian@ulster.ac.uk).}}

\newif\ifcompactlayout
\compactlayouttrue
\ifcompactlayout
\AtBeginDocument{%
  \setlength{\abovedisplayskip}{1pt plus 1pt minus 0.5pt}%
  \setlength{\belowdisplayskip}{1pt plus 1pt minus 0.5pt}%
  \setlength{\abovedisplayshortskip}{0pt plus 1pt}%
  \setlength{\belowdisplayshortskip}{1pt plus 1pt minus 0.5pt}%
  \setlength{\jot}{1pt}%
}
\fi

\allowdisplaybreaks[1]
\begin{document}
\maketitle

\begin{abstract}
This paper investigates dynamic channel tracking for high-rate Internet of things (IoT) in distributed massive multiple-input multiple-output (mMIMO) with low-resolution analog-to-digital converters (ADCs), where pilot and data activities are decoupled, pilots are reused, and mobile channels evolve across access slots. We hereby develop a quantization-aware joint coordinated Kalman filter (Q-JCKF) for known pilot activity and a quantization-aware distributed probabilistic data association filter (Q-DPDAF) for uncoordinated access. For uplink analysis, we derive a closed-form statistical use-and-then-forget (UatF) expression for Q-JCKF and mixture-aware, pilot-conditioned, and long-term statistical UatF formulations for Q-DPDAF, together with the large-N pilot-reuse ceiling and a cross-access point (AP) association term with an $\mathcal{O}(\epsilon)$ posterior-concentration bound. The proposed framework recovers JC-KF/PDAF in the ideal-ADC co-located limit, and quantifies the conditional loss of Kalman measurement information caused by finite-resolution quantization. We further characterize computational, memory, and fronthaul scaling: Kronecker-structured Q-JCKF reduces covariance-processing complexity, while groupwise Q-DPDAF limits hypothesis enumeration from $2^K$ to $2^{K_{r}}$ and enables parallel AP-local processing. Simulations corroborate the analytical and complexity outcomes, quantify ADC-resolution and pilot-access effects, and demonstrate improved lower-tail spectral efficiency, robust mobility tracking, near-ideal 4-bit performance, reliable Q-DPDAF activity inference, and substantial runtime reductions over legacy implementations and processing frameworks.
\end{abstract}


\section{Introduction}

\looseness=-1
Massive connectivity in fifth-generation (5G), beyond-5G, and emerging sixth-generation (6G) networks places stringent demands on channel state information (CSI) acquisition and uplink spectral efficiency \cite{BMZ24,GG21}. Although co-located massive multiple-input multiple-output (mMIMO) provides substantial spatial multiplexing gains, unequal large-scale fading limits its ability to maintain uniform service quality, particularly under imperfect power control. Distributed mMIMO addresses this limitation through geographically separated access points (APs), providing macro-diversity gains and improved lower-tail spectral efficiency \cite{WWYWCH13}. However, the resulting AP-dependent channel statistics complicate CSI acquisition and increase the processing and fronthaul requirements of cooperative reception \cite{ZDXC24}. Equipping distributed APs with complete radio-frequency (RF) chains and high-resolution analog-to-digital converters (ADCs) imposes considerable hardware and power costs \cite{CBF16,XSLLW23,AAL23}. While hybrid beamforming reduces RF-chain requirements, it introduces additional beam management and processing overhead \cite{AKSMKPM18}. Low-resolution ADCs provide a more scalable alternative while retaining per-antenna digital processing \cite{XSLLW23}. Their use, however, introduces signal-dependent quantization distortion whose statistics vary across APs with ADC resolution and aggregate received power \cite{AAL23}. Consequently, quantization-aware channel estimation must account for the aggregate pilot power at each AP, including contributions from other pilot groups that remain in the distortion covariance after despreading. In high-rate Internet of things (IoT) networks, frequent pilot transmissions under limited training resources lead to extensive pilot reuse, coupling the observations of pilot-sharing users and complicating channel estimation \cite{MAHM24}. Unknown pilot activity further couples activity detection with channel estimation \cite{RB23}, while mobility-induced channel variations require exploiting temporal correlation for recursive CSI tracking \cite{SF21,SF222}. When pilot and data activities are independent, such tracking is particularly important for maintaining CSI across slots in which data-active users do not transmit pilots. Previous studies \cite{SFR21,SF24,SF23,SF211,SF22} developed channel-tracking algorithms for high-rate IoT under pilot reuse and temporally correlated fading. Two pilot access scenarios were considered: coordinated access, where the base station knows the users' pilot transmission patterns, and uncoordinated access, where users select their pilot transmission slots without prior notification. For coordinated access, the joint coordinated Kalman filter (JC-KF) provides the minimum mean-square error (MMSE) estimate of the aggregate channel state while retaining the inter-user error cross-covariances induced by pilot collisions. For uncoordinated access, the probabilistic data association filter (PDAF) combines hypothesis-conditioned Kalman updates using posterior activity probabilities, retaining activity uncertainty through Gaussian-mixture moment matching. Although these algorithms establish a basis for channel tracking under pilot collisions, their extension to quantized distributed reception requires AP-specific channel and measurement statistics, activity-dependent quantization models, and likelihood fusion across geographically separated APs.

{\emph{ Related Works}}: Several massive random access studies have addressed activity detection and channel estimation in co-located mMIMO, often under simplified assumptions on user power control. Chen et al. \cite{CSY21} exploited large-scale fading differences across cells to improve sparse activity detection, while \cite{LY20} analyzed uplink spectral efficiency (SE) in industrial IoT and showed that maximum-ratio (MR) combining can outperform zero-forcing (ZF) under severe pilot interference. Moving toward distributed reception, \cite{YHMQJ20} investigated mixed-ADC mMIMO under different traffic conditions, employing compressed sensing for joint activity detection and channel estimation of light-load users and large-scale-fading-based dynamic clustering to support higher-throughput users. For asynchronous cell-free mMIMO, \cite{ZCCG25} formulated joint activity detection and channel estimation as a structured sparse-recovery problem and developed an asynchronous-aware matching pursuit algorithm. The authors of \cite{RRB22} exploited angular-domain block sparsity through centralized and decentralized sparse Bayesian learning, whereas \cite{SLGWZ26} employed covariance-based maximum-likelihood estimation and adaptive AP selection for joint activity and delay estimation. These approaches recover user activity and channel information from observations, but do not maintain temporally correlated channel estimates through recursive updates across successive access slots. Quantization-aware channel estimation and SE analysis have been investigated in cell-free mMIMO under different hardware and propagation models. The authors of \cite{XSLLW23} developed an LMMSE channel estimator for variable-resolution ADCs, characterized uplink SE under MR and MMSE combining, and investigated bit allocation across antennas and APs. Extending the hardware model, \cite{MLMYT23} considered finite-resolution ADCs and DACs over spatially correlated Rician fading channels, incorporating quantization effects into channel estimation and uplink processing. Related studies investigated rate-splitting multiple access with low-resolution ADCs \cite{AASAL24} and achievable SE under imperfect CSI \cite{NHXCI25}. While these studies characterize quantization effects on channel estimation and achievable rates, they do not incorporate activity-dependent quantization noise into recursive Bayesian channel updates or the likelihood evaluation of competing pilot-activity hypotheses. Massive grant-free access has also been investigated through sparse-signal recovery and statistical activity detection. The authors of \cite{BVPS17} compared compressed sensing with coded slotted ALOHA, examining detection performance and resource utilization, while \cite{MBD15} employed a Neyman--Pearson criterion for activity detection within a compressed sensing framework. In \cite{CXSA22}, angular-domain information is exploited to improve joint activity detection and channel estimation, whereas the recovery conditions of covariance-based activity detection were characterized in \cite{CSLY22} through phase-transition analysis. These methods exploit sparsity, spatial structure, or received-signal statistics to identify active devices. In high-rate IoT, however, frequent pilot transmissions can weaken the sparsity assumptions underlying several detection methods. More importantly, identifying active users in individual access slots does not provide the continuous CSI required when pilot and data activities are independent, particularly when data-active users transmit without pilots. {\emph{Main Contributions}}:  This paper develops quantization-aware channel tracking algorithms for coordinated and uncoordinated access in distributed mMIMO under pilot reuse, user mobility, uncertain pilot activity, and low-resolution ADCs. The main contributions cover channel tracking, uplink spectral-efficiency analysis, and implementation complexity, as follows:

\noindent\textbf{1)} A quantization-aware dynamic channel model is developed in distributed mMIMO with pilot reuse, mobility, and independent pilot and data activity. For coordinated access, Q-JCKF incorporates AP-dependent quantization distortion, preserves pilot-induced inter-user error cross-covariances, and admits an exact Kronecker reduction to the pilot-group user domain.

\noindent\textbf{2)} For uncoordinated access, Q-DPDAF combines AP-local hypothesis-conditioned Kalman updates with CPU-level activity-likelihood fusion. Groupwise hypothesis evaluation uses a prior-mean approximation of outside-group quantization power, while posterior-weighted Gaussian moment matching retains activity uncertainty without hard decisions.

\noindent\textbf{3)} A closed-form statistical use-and-then-forget (UatF) uplink signal-to-interference-plus-noise ratio (SINR) is derived for Q-JCKF, accounting for tracking uncertainty, data-phase quantization, multiuser interference, and coherent pilot reuse. The analysis establishes a finite SINR ceiling under persistent coherent same-pilot interference and linear SINR growth with the number of antennas per AP when this term vanishes.

\noindent\textbf{4)} For Q-DPDAF, mixture-aware pilot-conditioned, and long-term statistical UatF formulations are derived. The current-slot cross-AP association induced by the shared activity posterior is characterized, its contribution to effective-channel moments is quantified, and an \(\mathcal{O}(\epsilon)\) concentration bound is established together with the resulting rate impact. The analysis identifies when independent AP-local moment matching is accurate.

\noindent\textbf{5)} The proposed trackers are shown to recover the earlier JC-KF/PDAF framework \cite{SFR21} in the ideal-ADC, co-located limit. The conditional loss of Kalman measurement information due to finite-resolution quantization is quantified, together with the resulting estimation-error covariance ordering and its limitations for the final Q-DPDAF covariance and UatF rate.

\noindent\textbf{6)} The computational, memory, and fronthaul scaling of both trackers is characterized. The Q-JCKF Kronecker structure reduces covariance recursion from the \(NK_r\)-dimensional state domain to the \(K_r\)-dimensional user domain, while groupwise Q-DPDAF reduces hypothesis enumeration from \(2^K\) to \(2^{K_r}\) and supports parallel AP-local processing.

\noindent\textbf{7)} Numerical results validate the UatF analyses, cross-AP association bound, AP-local approximation, large-\(N\) pilot-reuse behavior, and complexity trends. They quantify the effects of distributed reception, ADC resolution, mobility, and pilot-access probability on spectral efficiency, tracking, and prior-aware activity inference, while confirming runtime gains over the corresponding legacy implementations.

\section{System Model and Problem Formulation}

\looseness=-1
This paper considers the uplink of a distributed mMIMO system with $L$ geographically separated APs, each equipped with $N$ antennas, jointly serving $K$ single-antenna IoT users. The APs are connected to a CPU via error-free fronthaul links. The system operates in time-division duplexing (TDD) mode, and each access slot contains $\tau_p$ pilot symbols followed by $\tau_u$ data symbols. Hence, $\tau_{\rm sl}=\tau_p+\tau_u$ and $\eta_u=\tau_u/\tau_{\rm sl}$. Let $\{\mathbf{\phi}_r\}_{r=1}^{\tau_p}$ denote $\tau_p$ orthogonal pilots of length $\tau_p$, normalized as
$\mathbf\phi_r^{\cH}\mathbf\phi_{r'}=\delta_{r,r'}$. We use constant-modulus pilots with $|[\mathbf{\phi}_r]_n|^2=1/\tau_p$, as provided, for example, by normalized DFT columns. Thus, $\sqrt{\tau_p\rho_p}\mathbf{\phi}_r$ has per-symbol pilot power $\rho_p$ and total energy $\tau_p\rho_p$. Because $K$ may exceed $\tau_p$, pilots are shared. The users are partitioned into disjoint groups $\{\mcG_r\}_{r=1}^{\tau_p}$, where all users in $\mcG_r$ employ $\mathbf{\phi}_r$. Let $K_r\triangleq|\mathcal G_r|$, so that $\sum_{r=1}^{\tau_p}K_r=K$. For a balanced assignment, $K_r=K/\tau_p$ whenever $K/\tau_p\in\mathbb N$. Let $q_t^{(k)},d_t^{(k)}\in\{0,1\}$ denote the pilot- and data-activity indicators of user $k$ in slot $t$. Define $\mathcal D_t\triangleq\{k:d_t^{(k)}=1\}$. Pilot activities are independent across users and slots, with $\Pr\{q_t^{(k)}=1\}=\lambda_k$. Pilot and data activities are also statistically independent; hence, a data-active user need not transmit a pilot in the same slot. The receiver therefore maintains channel estimates for all users across successive slots. Let $\mathbf p_t^{(k)}$ and $\mathbf c_l$ denote the position of user $k$ and AP $l$, respectively, giving $d_{l, t}^{(k)}=\|\mathbf p_t^{(k)}-\mathbf c_l\|$. The user position evolves as $\mathbf p_{t+1}^{(k)} = \mathbf p_t^{(k)} + T_s\mathbf v_t^{(k)}$, where $T_s$ is the interval between consecutive channel-state updates and $\mathbf v_t^{(k)}$ denotes the velocity vector of user $k$ during slot $t$. The large-scale fading coefficient $\beta_{l,t}^{(k)}$, determined by path loss and shadowing, is assumed known at each tracking update; uncertainty in its acquisition is not modeled. The channel from user $k$ to AP $l$ is $\mathbf g_{l,t}^{(k)} = \sqrt{\beta_{l,t}^{(k)}}\, \mathbf h_{l,t}^{(k)}$, where $\mathbf h_{l,t}^{(k)} \sim \CN(\mathbf 0,\mathbf I_N)$. To capture mobility-induced channel aging, we use a first-order Gauss--Markov model, $\mathbf h_{l,t}^{(k)} = \rho_t^{(k)} \mathbf h_{l,t-1}^{(k)} + \sqrt{1-|\rho_t^{(k)}| ^2}\, \mathbf e_{l,t}^{(k)}$,  where $\mathbf e_{l,t}^{(k)}\sim\CN(\mathbf 0,\mathbf I_N)$ is independent of the previous channel state. The fading processes are independent across users and APs, and the innovations are independent across slots. Following \cite{SFR21,SF21}, the one-step correlation coefficient $|\rho_t^{(k)}|\leq 1$ is obtained from the adopted Jakes correlation model. Combining small- and large-scale channel variations gives
\begin{equation}
\mathbf g_{l,t}^{(k)} = a_{l,t}^{(k)} \mathbf g_{l,t-1}^{(k)} + \mathbf u_{l,t}^{(k)}, 
\label{eq:mobility_channel_state} 
\end{equation}
where $a_{l,t}^{(k)} = \rho_t^{(k)} \sqrt{ \frac{\beta_{l,t}^{(k)}} {\beta_{l,t-1}^{(k)}} }$,  and $\mathbf u_{l,t}^{(k)} \sim \CN\!\big( \mathbf 0, \upsilon_{l,t}^{(k)}\mathbf I_N \big)$,  with $\upsilon_{l,t}^{(k)} \triangleq \beta_{l,t}^{(k)} \big(1-|\rho_t^{(k)}|^2\big)$. Since $|a_{l,t}^{(k)}|^2\beta_{l,t-1}^{(k)} +\upsilon_{l,t}^{(k)} =\beta_{l,t}^{(k)}$, the recursion preserves $\mathbf g_{l,t}^{(k)} \sim\CN(\mathbf 0,\beta_{l,t}^{(k)}\mathbf I_N)$ conditioned on the large-scale fading trajectory. The channel is assumed constant over the pilot and data phases of each access slot and evolves across slots according to \eqref{eq:mobility_channel_state}.

\subsection{Pilot Observation Before and After Quantization}

\looseness=-1
During the pilot phase, users with $q_t^{(k)}=1$ transmit their assigned pilots. Before quantization, AP $l$ receives
\begin{equation}
\mathbf Y_{l,t}^{p} = \sqrt{\tau_p\rho_p} \sum_{r=1}^{\tau_p} \sum_{k\in\mcG_r} q_t^{(k)} \mathbf g_{l,t}^{(k)} \mathbf{\phi}_r^{\cH} + \mathbf W_{l,t}^{p}, \label{eq:pilot_pre_adc}
\end{equation}
where $\mathbf W_{l,t}^{p}\in\mathbb C^{N\times\tau_p}$ has i.i.d. $\CN(0,1)$ entries, so the thermal-noise variance is normalized to unity. The noise matrices are independent across APs and slots and independent of all channel processes. Each antenna at AP $l$ uses a $b_l$-bit ADC. The Gaussian additive quantization noise model (AQNM) yields $\widetilde{\mathbf Y}_{l,t}^{p} = \alpha_l\mathbf Y_{l,t}^{p} + \mathbf Z_{l,t}^{p}$, where $0<\alpha_l\leq1$ is the quantization gain and $\mathbf Z_{l,t}^{p}$ is the distortion term. At sufficiently high resolution, $\alpha_l \approx 1-\frac{\pi\sqrt{3}} {2}2^{-2b_l}$.  

\looseness=-1
The AQNM is used only as a Gaussian second-order surrogate for the nonlinear quantizer in which $\mathbf Z_{l,t}^{p}$ is modeled as a zero-mean proper Gaussian distortion that is uncorrelated with the ADC input, white across pilot samples, and independent across APs. Despreading with $\mathbf{\phi}_r$ gives
\begin{align}
\widetilde{\mathbf y}_{lr,t}^{p} &\triangleq \widetilde{\mathbf Y}_{l,t}^{p}\mathbf{\phi}_r \nonumber\\
&= \alpha_l\sqrt{\tau_p\rho_p} \sum_{k\in\mcG_r} q_t^{(k)}\mathbf g_{l,t}^{(k)} +
\alpha_l\mathbf W_{l,t}^{p}\mathbf{\phi}_r + \mathbf Z_{l,t}^{p}\mathbf{\phi}_r .
\label{eq:despread_expanded}
\end{align}
Define $ \kappa_l \triangleq \alpha_l\sqrt{\tau_p\rho_p}$ and $\mathbf n_{lr,t} \triangleq
\alpha_l\mathbf W_{l,t}^{p}\mathbf{\phi}_r + \mathbf Z_{l,t}^{p}\mathbf{\phi}_r$. Then $\widetilde{\mathbf y}_{lr,t}^{p} = \kappa_l \sum_{k\in\mcG_r} q_t^{(k)} \mathbf g_{l,t}^{(k)} + \mathbf n_{lr,t}$. 

\looseness=-1
 Let $\mathbf{q}_t=[q_t^{(1)},\ldots,q_t^{(K)}]^T$, and let $\mathbf{y}_{l,t}^{p}[n]$ and $\mathbf{z}_{l,t}^{p}[n]$
denote the $n$th columns of $\mathbf{Y}_{l,t}^{p}$ and $\mathbf{Z}_{l,t}^{p}$, respectively. For the $n$th pilot symbol, the constant-modulus property gives
\begin{equation}
\mathbb{E}\!\Big[ \mathbf{y}_{l,t}^{p}[n] \big(\mathbf{y}_{l,t}^{p}[n]\big)^{\cH} \mid \mathbf{q}_t \Big] = \Big( 1+\rho_p\sum_{k=1}^{K} q_t^{(k)}\beta_{l,t}^{(k)} \Big)\mathbf{I}_N .
\end{equation}
This covariance averages over small-scale fading and receiver noise while conditioning on the large-scale coefficients and pilot activity.

\looseness=-1
Moreover, under the adopted temporally white Gaussian-AQNM approximation, the distortion is modeled as uncorrelated across pilot samples and independent across APs, with
$\mathbb E\!\Big[ \mathbf z_{l,t}^{p}[n] \big(\mathbf z_{l,t}^{p}[m]\big)^{\cH} \mid \mathbf q_t \Big] \; \quad= \delta_{n,m}\alpha_l(1-\alpha_l) \Big( 1+ \rho_p \sum_{k=1}^{K} q_t^{(k)} \beta_{l,t}^{(k)} \Big) \mathbf I_N.$
Since $\|\mathbf{\phi}_r\|^2=1$, despreading preserves this distortion covariance, whereas the filtered thermal-noise covariance is $\alpha_l^2\mathbf I_N$. Therefore, $\mathbb E\!\big[ \mathbf n_{lr,t}\mathbf n_{lr,t}^{\cH} \mid \mathbf q_t \big]=
\sigma_{l,t}^{2}(\mathbf q_t)\mathbf I_N$, where
\begin{equation}
\sigma_{l,t}^{2}(\mathbf q_t) = \alpha_l \big[ 1+ (1-\alpha_l)\rho_p \sum_{k=1}^{K} q_t^{(k)} \beta_{l,t}^{(k)} \big]. \label{eq:sigma_coord}
\end{equation}
Accordingly, $\mathbf n_{lr,t}\mid\mathbf q_t \sim \CN\!\big( \mathbf 0, \sigma_{l,t}^{2}(\mathbf q_t)\mathbf I_N \big)$. Although despreading removes the signals from other pilot groups, their received power still contributes to quantization noise. For unknown activity, Section~III-B approximates this outside-group contribution by its prior mean. For unknown activity, Section~III-B replaces the outside-group contribution by its prior mean.

\looseness=-1
Finally, for $\mcG_r=\{k_1,\ldots,k_{K_r}\}$, define $\mathbf q_{r,t} \triangleq [q_t^{(k_1)},\ldots,q_t^{(k_{K_r})}]^{\cT}$, and the stacked state $\mathbf x_{lr,t} \triangleq \left[ (\mathbf g_{l,t}^{(k_1)})^{\cT}, \ldots, (\mathbf g_{l,t}^{(k_{K_r})})^{\cT} \right]^{\cT}$. The group dynamics are
\begin{equation}
\mathbf x_{lr,t} = \mathbf A_{lr,t}\mathbf x_{lr,t-1} + \mathbf u_{lr,t}, \label{eq:group_state_transition}
\end{equation}
where $\mathbf A_{lr,t} = \operatorname{blkdiag}\!\big( a_{l,t}^{(k_1)}\mathbf I_N, \ldots, a_{l,t}^{(k_{K_r})}\mathbf I_N \big)$, $\mathbf u_{lr,t} \sim \CN(\mathbf 0,\mathbf Q_{lr,t})$, and  $\mathbf Q_{lr,t} = \operatorname{blkdiag}\!\big( \upsilon_{l,t}^{(k_1)}\mathbf I_N, \ldots, \upsilon_{l,t}^{(k_{K_r})}\mathbf I_N \big)$.  
The corresponding observation model is
\begin{equation}
\widetilde{\mathbf y}_{lr,t}^{p}= \mathbf H_{lr}(\mathbf q_{r,t}) \mathbf x_{lr,t} +
\mathbf n_{lr,t}, \label{eq:group_state_observation}
\end{equation}
where $\mathbf H_{lr}(\mathbf q_{r,t}) = \kappa_l \big( \mathbf q_{r,t}^{\cT} \otimes \mathbf I_N \big)$.  Equations~\eqref{eq:group_state_transition} and \eqref{eq:group_state_observation} define the quantization-aware state-space model used for channel tracking at all APs. Pilot activity is known in coordinated access, whereas uncoordinated access requires joint inference of activity and channel states from distributed observations.

\section{Quantization-Aware Channel Tracking}

\looseness=-1
Based on the groupwise state-space model of Section~II, we develop two quantization-aware recursive trackers. Q-JCKF assumes the full pilot-activity vector $\mathbf q_t$ is known and admits an exact user-domain covariance reduction. Q-DPDAF treats the activity pattern as unknown and combines AP-local conditional Kalman updates with CPU likelihood fusion and Gaussian-mixture moment matching. Section~III-C relates both trackers to the ideal-ADC JC-KF/PDAF framework.

\subsection{Coordinated Q-JCKF}

\looseness=-1
In the coordinated case, both $\mathbf H_{lr}(\mathbf q_{r,t})$ in \eqref{eq:group_state_observation} and $\sigma_{l,t}^{2}(\mathbf q_t)$ in \eqref{eq:sigma_coord} are available. Under the Gaussian AQNM surrogate, the resulting state-space model is linear Gaussian; hence, Q-JCKF provides the conditional MMSE estimate. Define $
\mathbf D_{lr,t} \triangleq \diag\!\big( a_{l,t}^{(k_1)},\ldots,a_{l,t}^{(k_{K_r})} \big)$ and $\mathbf V_{lr,t} \triangleq \diag\!\big( \upsilon_{l,t}^{(k_1)},\ldots, \upsilon_{l,t}^{(k_{K_r})} \big)$, so that $\mathbf A_{lr,t}=\mathbf D_{lr,t}\otimes\mathbf I_N$ and $\mathbf Q_{lr,t}=\mathbf V_{lr,t}\otimes\mathbf I_N$. Given the posterior moments $\mathbf m_{lr,t-1}^{+}$ and $\mathbf P_{lr,t-1}^{+}$, the one-step predictions are
$\mathbf m_{lr,t}^{-} = \mathbf A_{lr,t}\mathbf m_{lr,t-1}^{+}, \; \mathbf P_{lr,t}^{-} = \mathbf A_{lr,t}\mathbf P_{lr,t-1}^{+} \mathbf A_{lr,t}^{\cH} + \mathbf Q_{lr,t}.$

\looseness=-1
Let $\mathbf H_{lr,t}\triangleq \mathbf H_{lr}(\mathbf q_{r,t})$. The innovation covariance and Kalman gain are $\mathbf S_{lr,t} = \mathbf H_{lr,t}\mathbf P_{lr,t}^{-} \mathbf H_{lr,t}^{\cH} + \sigma_{l,t}^{2}(\mathbf q_t)\mathbf I_N$ and $\mathbf K_{lr,t}= \mathbf P_{lr,t}^{-}\mathbf H_{lr,t}^{\cH} \mathbf S_{lr,t}^{-1}$. The posterior update is
$\mathbf m_{lr,t}^{+}= \mathbf m_{lr,t}^{-} + \mathbf K_{lr,t} \big( \widetilde{\mathbf y}_{lr,t}^{p} - \mathbf H_{lr,t}\mathbf m_{lr,t}^{-} \big), \; \mathbf P_{lr,t}^{+} = \mathbf P_{lr,t}^{-} - \mathbf K_{lr,t}\mathbf S_{lr,t} \mathbf K_{lr,t}^{\cH},$
where $\mathbf m_{lr,t}^{\pm}\in\mathbb C^{NK_r}$ and $\mathbf P_{lr,t}^{\pm}\in
\mathbb C^{NK_r\times NK_r}$. Conditioned on the known pilot activity and slow-time channel statistics, these recursions yield the MMSE estimate within the adopted linear-Gaussian AQNM model; this optimality does not apply to the underlying nonlinear quantizer.

\looseness=-1
Let $\mathbf{P}_{lr,t}^{\pm}$ denote the full predicted/posterior state covariance. For initialization, the channel prior in Section~II gives $\mathbf m_{lr,0}^{+}=\mathbf {0}_{NK_{r}}$, $\overline{\mathbf P}_{lr,0}^{+} =\operatorname{diag}\!\big( \beta_{l,0}^{(k_1)},\ldots, \beta_{l,0}^{(k_{K_r})} \big)$, and $\mathbf P_{lr,0}^{+} = \overline{\mathbf P}_{lr,0}^{+}\otimes\mathbf I_N$ with $\overline{\mathbf{P}}_{lr,0}^{+}\in  \mathbb{C}^{K_r\times K_r}$. The state dynamics, process noise, and observation model preserve the structure $\mathbf{P}_{lr,t}^{\pm} =\overline{\mathbf{P}}_{lr,t}^{\pm} \otimes\mathbf{I}_N$, where $\overline{\mathbf{P}}_{lr,t}^{\pm}$ is the corresponding user-domain covariance. Its prediction is therefore
\begin{equation}
\overline{\mathbf{P}}_{lr,t}^{-} = \mathbf{D}_{lr,t} \overline{\mathbf{P}}_{lr,t-1}^{+} \mathbf{D}_{lr,t}^{\cH} + \mathbf{V}_{lr,t}. \label{eq:reduced_prediction}
\end{equation}

\looseness=-1
For the known group-activity vector $\mathbf q_{r,t}$, the innovation covariance reduces to $\mathbf S_{lr,t}=s_{lr,t}\mathbf I_N$, where  $s_{lr,t} = \kappa_l^2 \mathbf q_{r,t}^{T} \overline{\mathbf P}_{lr,t}^{-} \mathbf q_{r,t} + \sigma_{l,t}^{2}(\mathbf q_t)$. The dependence of $\sigma_{l,t}^{2}(\mathbf q_t)$ on the full network pilot-activity vector $\mathbf q_t$ results from quantization before pilot despreading. Accordingly, $\mathbf{K}_{lr,t} = \frac{\kappa_l}{s_{lr,t}} \big( \overline{\mathbf{P}}_{lr,t}^{-} \mathbf{q}_{r,t} \otimes\mathbf{I}_N \big)$, and the posterior user-domain covariance becomes
\begin{equation}
\overline{\mathbf{P}}_{lr,t}^{+} = \overline{\mathbf{P}}_{lr,t}^{-} - \frac{\kappa_l^2}{s_{lr,t}} \overline{\mathbf{P}}_{lr,t}^{-} \mathbf{q}_{r,t}\mathbf{q}_{r,t}^{T} \overline{\mathbf{P}}_{lr,t}^{-}. \label{eq:reduced_update}
\end{equation}
Hence, although the channel state has dimension $NK_r$, the Q-JCKF covariance recursion operates in the $K_r$-dimensional user domain.

\looseness=-1
Equations~\eqref{eq:reduced_prediction} and \eqref{eq:reduced_update} also show how pilot reuse creates and propagates cross-covariances between users in the same pilot group. In particular, when multiple users in the group are pilot-active, their common observation can generate nonzero off-diagonal entries in $\overline{\mathbf P}_{lr,t}^{+}$; previously generated cross-covariances are subsequently propagated by
\eqref{eq:reduced_prediction}. These off-diagonal terms determine the coherent same-pilot interference in the Q-JCKF UatF analysis of Section~IV-A.

\subsection{Uncoordinated Q-DPDAF}

\looseness=-1
With unknown pilot activity, both the measurement matrix and the AQNM noise variance depend on the activity hypothesis. Enumerating all $2^K$ activity patterns is impractical for massive random access. Instead, Q-DPDAF considers the $2^{K_r}$ patterns within each pilot-sharing group and exploits the observations collected by all APs. For $\mcG_r=\{k_1,\ldots,k_{K_r}\}$, define the unknown group activity vector as $\mathbf q_{r,t}=[q_t^{(k_1)},\ldots,q_t^{(k_{K_r})}]^{\cT}$. The Q-DPDAF evaluates candidate vectors $\mathbf q\in\mcQ_r$, where $\mcQ_r\triangleq\{0,1\}^{K_r}$. Assuming independent Bernoulli pilot access with probabilities $\{\lambda_k\}_{k=1}^{K}$, the prior probability of $\bm q$ is $\pi_r(\mathbf q) = \prod_{j=1}^{K_r} \lambda_{k_j}^{q^{(j)}} (1-\lambda_{k_j})^{1-q^{(j)}}$. Because quantization precedes despreading, each hypothesis affects both the useful pilot signal and the noise variance, which also depends on users outside $\mcG_r$. To avoid enumerating their activities, we replace their received pilot power by its prior mean, $\bar\chi_{l,-r,t} \triangleq \E\!\big[ \sum_{j\notin\mcG_r} q_t^{(j)}\beta_{l,t}^{(j)} \big] = \sum_{j\notin\mcG_r} \lambda_j\beta_{l,t}^{(j)}$. The resulting hypothesis-dependent variance is
\begin{equation}
\widehat{\sigma}_{lr,t}^{2}(\mathbf q) = \alpha_l  \Big[1+ (1-\alpha_l)\rho_p \big(
\sum_{j=1}^{K_r} q^{(j)} \beta_{l,t}^{(k_j)} + \bar\chi_{l,-r,t} \big) \Big].
\label{eq:dpdaf_meanfield_variance}
\end{equation}
Unlike the coordinated variance in \eqref{eq:sigma_coord}, this mean-field approximation retains the candidate-dependent in-group power but not the instantaneous outside-group fluctuations.

\looseness=-1
For $\mathbf q\in\mcQ_r$, the measurement matrix is $\mathbf H_{lr}(\mathbf q)=\kappa_l(\mathbf q^{\cT}\otimes\I_N)$ and  the effective noise is $\mathbf n_{lr,t}\mid\mathbf q\sim\CN(\zero,\widehat{\sigma}_{lr,t}^{2}(\mathbf q)\I_N)$. Each hypothesis thus defines a linear-Gaussian observation model. All activity hypotheses at a given AP start from the same predicted state $(\mathbf m_{lr,t}^{-},\mathbf P_{lr,t}^{-})$; they differ only through their measurement matrix and effective AQNM noise variance. With $\mathbf e_{lr,t}(\mathbf q) =\widetilde{\mathbf y}_{lr,t}^{p}-\mathbf H_{lr}(\mathbf q)\mathbf m_{lr,t}^{-}$,
\begin{align*}
& \mathbf S_{lr,t}(\mathbf q) =\mathbf H_{lr}(\mathbf q)\mathbf P_{lr,t}^{-}\mathbf H_{lr}(\mathbf q)^{\cH} +\widehat{\sigma}_{lr,t}^{2}(\mathbf q)\I_N,\\
& \mathbf K_{lr,t}(\mathbf q) =\mathbf P_{lr,t}^{-}\mathbf H_{lr}(\mathbf q)^{\cH}
\mathbf S_{lr,t}(\mathbf q)^{-1},
\end{align*}
the conditional Kalman updates are
$\mathbf m_{lr,t}^{+}(\mathbf q) =\mathbf m_{lr,t}^{-}+\mathbf K_{lr,t}(\mathbf q)\mathbf e_{lr,t}(\mathbf q), \; \mathbf P_{lr,t}^{+}(\mathbf q) =\mathbf P_{lr,t}^{-}-\mathbf K_{lr,t}(\mathbf q)\mathbf S_{lr,t}(\mathbf q) \mathbf K_{lr,t}(\mathbf q)^{\cH}.$

\looseness=-1
Let $\mcY_{t-1}$ denote all pilot observations through slot $t-1$. To combine AP observations, the Q-DPDAF uses the factorized predicted-state approximation
\begin{align}
p(\mathbf x_{1r,t},\ldots,\mathbf x_{Lr,t}\mid\mcY_{t-1})\approx \prod_{l=1}^{L}\CN\big(\mathbf x_{lr,t}; \mathbf m_{lr,t}^{-},\mathbf P_{lr,t}^{-}\big). \label{eq:dpdaf_factorized_prior}
\end{align}
Hence, cross-AP dependence inherited from earlier mixture updates is not propagated. With independent AP noise, the predictive likelihood factorizes as $p(\{\widetilde{\mathbf y}_{lr,t}^{p}\}_{l=1}^{L}\mid\mathbf q,\mcY_{t-1})\approx\prod_{l=1}^{L}\mathcal L_{lr,t}(\mathbf q)$, where marginalizing the Gaussian predicted state gives $\mathcal L_{lr,t}(\mathbf q)=\CN\big(\widetilde{\mathbf y}_{lr,t}^{p};\mathbf H_{lr}(\mathbf q)\mathbf m_{lr,t}^{-},\mathbf S_{lr,t}(\mathbf q)\big)$. Its log-likelihood is
$\log\mathcal L_{lr,t}(\mathbf q) =-N\log\pi-\log\det\mathbf S_{lr,t}(\mathbf q)\; \quad- \big(\mathbf e_{lr,t}(\mathbf q)\big)^{\cH} \big(\mathbf S_{lr,t}(\mathbf q)\big)^{-1}\mathbf e_{lr,t}(\mathbf q).$

\looseness=-1
Each AP sends one log-likelihood per hypothesis to the CPU, without forwarding raw pilot observations or conditional covariances. The CPU forms $\ell_{r,t}(\mathbf q)=\log\pi_r(\mathbf q)+\sum_{l=1}^{L}\log\mathcal L_{lr,t}(\mathbf q)$ and computes the common posterior weights
$\omega_{r,t}(\mathbf q)= \frac{\exp\big(\ell_{r,t}(\mathbf q)-\ell_{r,t}^{\max}\big)} {\displaystyle\sum_{\mathbf q'\in\mcQ_r} \exp\big(\ell_{r,t}(\mathbf q')-\ell_{r,t}^{\max}\big)},$
where $\ell_{r,t}^{\max}=\max_{\mathbf q\in\mcQ_r}\ell_{r,t}(\mathbf q)$ prevents numerical overflow. The CPU returns the posterior weights to the APs for local mixture moment matching. If needed, activity is detected by $\widehat{\mathbf q}_{r,t}^{\mathrm{MAP}} =\arg\max_{\mathbf q\in\mcQ_r}\omega_{r,t}(\mathbf q)$. This hard decision is not used by the channel tracker, which instead retains all posterior weights to preserve pilot-activity uncertainty.

\looseness=-1
The conditional updates and shared weights define the current-slot joint posterior approximation
\begin{align}
&p(\mathbf x_{1r,t},\ldots,\mathbf x_{Lr,t}\mid\mcY_t)\nonumber\\
&\quad\approx\sum_{\mathbf q\in\mcQ_r}\omega_{r,t}(\mathbf q) \prod_{l=1}^{L}\CN\big(\mathbf x_{lr,t};\mathbf m_{lr,t}^{+}(\mathbf q),\mathbf P_{lr,t}^{+}(\mathbf q)\big),
\label{eq:mixture}
\end{align}
where $\mcY_t$ includes the pilot observations through slot $t$. In this approximation, AP states are conditionally independent given $\mathbf q$, but mostly dependent after marginalizing the shared hypothesis.

\looseness=-1
To avoid exponential growth in the number of mixture components across slots, each AP replaces its marginal mixture by a Gaussian with matching the first and second moments. Define $\mathbf m_{lr,t}^{\mathrm{MM}} \triangleq \sum_{\mathbf q\in\mathcal Q_r} 
\omega_{r,t}(\mathbf q) \mathbf m_{lr,t}^{+}(\mathbf q)$ and $\mathbf{\Delta}_{lr,t}(\mathbf q) \triangleq \mathbf m_{lr,t}^{+}(\mathbf q) - \mathbf m_{lr,t}^{\mathrm{MM}}$.  The corresponding moment-matched covariance is
\begin{align} 
\mathbf P_{lr,t}^{\mathrm{MM}} &\triangleq \sum_{\mathbf q\in\mathcal Q_r} \omega_{r,t}(\mathbf q) \Big[ \mathbf P_{lr,t}^{+}(\mathbf q) + \mathbf{\Delta}_{lr,t}(\mathbf q)\big(\mathbf{\Delta}_{lr,t}(\mathbf q)\big)^{\cH} \Big]. \label{eq:dpdaf_mm_cov_expanded}
\end{align}

\looseness=-1
The two terms in \eqref{eq:dpdaf_mm_cov_expanded} represent conditional estimation uncertainty and uncertainty across activity hypotheses; i.e., omitting the latter would underestimate the mixture covariance. The tracker sets $\mathbf m_{lr,t}^{+}=\mathbf m_{lr,t}^{\mathrm{MM}} $ and $\mathbf P_{lr,t}^{+}=\mathbf P_{lr,t}^{\mathrm{MM}}$ for the next prediction. The between-hypothesis outer-product term generally destroys the Kronecker structure $\overline{\mathbf P}\otimes\mathbf I_N$. Consequently, Q-DPDAF propagates a full $NK_r\times NK_r$ AP-local covariance, unlike Q-JCKF. The shared activity hypothesis also induces a current-slot cross-AP posterior association covariance, characterized in Section~IV-C. This dependence is retained for current-slot rate evaluation but is not propagated as a joint cross-AP covariance to the next slot.

\begin{remark}
Q-DPDAF relies on four approximations: (i) the Gaussian AQNM for nonlinear ADCs, (ii) the prior-mean approximation of outside-group pilot power in \eqref{eq:dpdaf_meanfield_variance}, (iii) the factorized AP-local predicted posterior in \eqref{eq:dpdaf_factorized_prior}, which neglects cross-AP dependence inherited from previous slots, and (iv) Gaussian moment matching in \eqref{eq:dpdaf_mm_cov_expanded} for subsequent prediction. Conditioned on each activity hypothesis, the AP-local Kalman update is exact under the adopted linear-Gaussian surrogate, while \eqref{eq:dpdaf_mm_cov_expanded} preserves the exact first two moments of the AP-local mixture. Thus, global MMSE optimality is not claimed for the original nonlinear quantized system.
\end{remark}

\subsection{Reduction to JC-KF/PDAF and Quantization Penalty}

\looseness=-1
 As $\alpha_l\rightarrow1$, the AQNM distortion vanishes, $\kappa_l\rightarrow \sqrt{\tau_p\rho_p}$, and the effective pilot-noise covariance approaches the ideal-ADC value. Hence, Q-JCKF and Q-DPDAF reduce to their ideal-ADC counterparts; the legacy JC-KF/PDAF model in \cite{SFR21} is recovered for $L=1$, $N=M$, $K_r=K$, and otherwise matched model parameters. To quantify the finite-resolution penalty, consider the same geometry, pilot activity, pilot energy, and predicted covariance in the ideal- and quantized-ADC models. For the coordinated case, define $B_{l,t}^{p} \triangleq \sum_{k=1}^{K}q_t^{(k)} \beta_{l,t}^{(k)}$ and $\mathbf U_{r,t} \triangleq (\mathbf q_{r,t}\mathbf q_{r,t}^{T}) \otimes\mathbf I_N$. The measurement-information increment $\mathbf J=\mathbf H^{\cH}\mathbf R^{-1}\mathbf H$ is then $\mathbf J_{lr,t}^{\rm ideal} = \tau_p\rho_p\mathbf U_{r,t}$ and $\mathbf J_{lr,t}^{Q} =\vartheta_{l,t} \mathbf J_{lr,t}^{\rm ideal}$, where $\vartheta_{l,t} = \frac{\alpha_l}{1+(1-\alpha_l)\rho_p B_{l,t}^{p}}$  and $0<\vartheta_{l,t}\leq1$. Hence, $\mathbf J_{lr,t}^{Q} \preceq \mathbf J_{lr,t}^{\rm ideal}$. Since $\mathbf{P}^{+}=[(\mathbf{P}^{-})^{-1} +\mathbf{J}]^{-1} $,  a common positive-definite predicted covariance gives $\mathbf P_{lr,Q,t}^{+} \succeq \mathbf P_{lr, \mathrm{ideal},t}^{+}$. Thus, under the adopted Gaussian AQNM surrogate, quantization cannot improve the conditional Kalman error covariance. For the coordinated tracker, this ordering is preserved recursively when the two systems use the same activity sequence, state transitions, process-noise covariances, and initial statistics. For Q-DPDAF, with a common predicted covariance, the same information ordering holds hypothesis-wise after replacing $B_{l,t}^{p}$ by $\widehat{B}_{lr,t}^{p}(\mathbf q)
= \sum_{j=1}^{K_r} q^{(j)}\beta_{l,t}^{(k_j)} + \bar\chi_{l,-r,t}$. This ordering does not generally extend recursively to the final moment-matched covariance, since ADC resolution also affects the hypothesis likelihoods and, as a result, the posterior weights, conditional means, and between-hypothesis covariance. A covariance ordering also does not imply a universal UatF-rate ordering because the rate depends on interference and data-phase quantization, as well.

\section{Uplink Spectral-Efficiency Analysis}

\looseness=-1
We derive uplink UatF spectral-efficiency bounds for local maximum-ratio combining (MRC) followed by coherent CPU combining, accounting for tracking uncertainty, pilot reuse, multiuser interference, and data-phase ADC quantization. Since pilot and data activities are independent, a data-active user need not transmit a pilot in the current slot; its recursively maintained channel estimate remains available for data detection. We condition on the data-active set $\mathcal D_t$ and the known large-scale fading, thus excluding data-activity detection errors; these variables are omitted from the conditioning notation below. Before quantization, AP $l$ receives $\mathbf y_{l,t}^{u}=\sqrt{\rho_u} \sum_{i\in\mcD_t} \mathbf g_{l,t}^{(i)} s_t^{(i)}+\mathbf w_{l,t}^{u}$, where $\rho_u$ is the normalized data transmit power, $\{s_t^{(i)}\}_{i\in\mathcal D_t}$ are independent $\CN(0,1)$ symbols independent of the channels and noise, and $\mathbf w_{l,t}^{u}\sim\CN(\mathbf 0,\mathbf I_N)$. At each AP, the same ADC resolution is used during the pilot and data phases. Applying the Gaussian AQNM of Section~II-A gives $\widetilde{\mathbf y}_{l,t}^{u}=\alpha_l\sqrt{\rho_u}\sum_{i\in\mcD_t}\mathbf g_{l,t}^{(i)} s_t^{(i)} +\alpha_l\mathbf w_{l,t}^{u}+\mathbf z_{l,t}^{u}$. Let $B_{l,t}\triangleq\sum_{i\in\mcD_t}\beta_{l,t}^{(i)}$ be the aggregate large-scale received power of data-active users. Spatially independent Rayleigh fading and independent unit-power symbols yield $\E\!\big[\mathbf y_{l,t}^{u}(\mathbf y_{l,t}^{u})^{\cH} \mid\mcD_t\big]=(1+\rho_u B_{l,t})\mathbf I_N$. The AQNM distortion covariance is thus
\begin{equation}
\E\!\big[ \mathbf z_{l,t}^{u} (\mathbf z_{l,t}^{u})^{\cH} \mid\mathcal D_t \big] =
\alpha_l(1-\alpha_l) (1+\rho_u B_{l,t})\mathbf I_N. \label{eq:data_quant_noise_cov}
\end{equation}
Unlike \eqref{eq:sigma_coord}, this data-phase covariance depends on $\mcD_t$ and is averaged over data symbols and small-scale fading rather than conditioned on instantaneous channels.

\looseness=-1
For the rate analysis, $\mathbf z_{l,t}^{u}\mid\mcD_t$ is modeled as a proper Gaussian with covariance \eqref{eq:data_quant_noise_cov}, uncorrelated with the ADC input, and independent of the instantaneous data symbols, thermal noise, pilot observations, and distortions at other APs. This Gaussian-AQNM model is a second-order surrogate rather than an exact sample-wise model of the nonlinear quantizer. With $\mathbf n_{l,t}^{u}\triangleq\alpha_l\mathbf w_{l,t}^{u}+\mathbf z_{l,t}^{u}$, its covariance and the compact received-signal model become
\begin{align}
\mathbf R_{n,l,t}^{u} &\triangleq \E[\mathbf n_{l,t}^{u}(\mathbf n_{l,t}^{u})^{\cH} 
\mid\mcD_t] \nonumber\\
&= \alpha_l[1+(1-\alpha_l)\rho_u B_{l,t}] \mathbf I_N. \label{eq:data_effective_noise_cov}
\end{align}
$\widetilde{\mathbf y}_{l,t}^{u} =\alpha_l\sqrt{\rho_u}\sum_{i\in\mcD_t} \mathbf g_{l,t}^{(i)}s_t^{(i)}+\mathbf n_{l,t}^{u}.$

\looseness=-1
For user $k$, AP $l$ uses $\mathbf v_{l,t}^{(k)}=\widehat{\mathbf g}_{l,t}^{(k)}\in\mathbb C^N$, obtained from the Q-JCKF or Q-DPDAF, and forwards $(\mathbf v_{l,t}^{(k)})^{\cH}\widetilde{\mathbf y}_{l,t}^{u}$ to the CPU. The decision statistic $\widehat s_t^{(k)}=\sum_{l=1}^{L} (\mathbf v_{l,t}^{(k)})^{\cH}\widetilde{\mathbf y}_{l,t}^{u}$ is
\begin{align}
\widehat s_t^{(k)} &=\sqrt{\rho_u}\,h_t^{(k,k)}s_t^{(k)} +\sqrt{\rho_u} \sum_{\substack{i\in\mcD_t\\i\ne k}}  h_t^{(k,i)}s_t^{(i)}+\widetilde n_t^{(k)},
\label{eq:cpu_signal_decomposition}
\end{align}
where $\widetilde n_t^{(k)}\triangleq\sum_l(\mathbf v_{l,t}^{(k)})^{\cH}\mathbf n_{l,t}^{u}$ and
\begin{equation}
h_t^{(k,i)}\triangleq\sum_{l=1}^{L} \alpha_l(\mathbf v_{l,t}^{(k)})^{\cH}\mathbf g_{l,t}^{(i)}. \label{eq:effective_combined_channel}
\end{equation}

\looseness=-1
Let $\mathcal I_t$ denote the conditioning information used in the UatF analysis, including $\mathcal D_t$, the known large-scale fading, and, when applicable, pilot observations or posterior statistics. Define $\mu_t^{(k)} \triangleq \E[h_t^{(k,k)}\mid\mathcal I_t]$ and $\Psi_t^{(k,i)} \triangleq \E[|h_t^{(k,i)}|^2\mid\mathcal I_t]$. Using \eqref{eq:data_effective_noise_cov} and the independence of the AP noises, the conditional mean noise power is
$\bar\nu_t^{(k)}(\mathcal I_t) = \sum_{l=1}^{L} \alpha_l[1+(1-\alpha_l)\rho_u B_{l,t}] \E\!\left[ \|\mathbf v_{l,t}^{(k)}\|^2 \mid\mathcal I_t \right].$

\looseness=-1
The desired-channel fluctuation $h_t^{(k,k)}-\mu_t^{(k)}$ has zero conditional mean and power $\Psi_t^{(k,k)}-|\mu_t^{(k)}|^2$; under the Gaussian-AQNM surrogate and independent Gaussian signaling, it combines with multiuser interference and effective noise as an uncorrelated disturbance. Hence, for $k\in\mcD_t$,
\begin{equation}
\mathrm{SINR}_t^{(k)}(\mcI_t) =\frac{\rho_u|\mu_t^{(k)}|^2}{\rho_u \displaystyle \sum_{i\in\mcD_t}\Psi_t^{(k,i)} -\rho_u|\mu_t^{(k)}|^2+\bar\nu_t^{(k)}(\mcI_t)}.
\label{eq:uatf_general}
\end{equation}
If all combining vectors vanish, the SINR is defined as zero. The Gaussian-signaling worst-case uncorrelated-noise argument under the adopted AQNM model gives the conditional bound $R_t^{(k)}(\mcI_t)=\eta_u\log_2\!\big( 1+\mathrm{SINR}_t^{(k)}(\mcI_t)\big)$.

\looseness=-1
The generic UatF bound is specialized differently for the two trackers. For Q-JCKF, the moments are averaged over pilot observations and small-scale fading conditioned on the known activity history, yielding the statistical UatF bound in Section~IV-A. For Q-DPDAF, conditioning on the pilot observations fixes the mixture-based combining vectors, leading to the pilot-conditioned bound in Section~IV-B. Averaging the underlying moments before forming the SINR is generally not equivalent to averaging pilot-conditioned rates.

\subsection{Q-JCKF Statistical UatF Analysis}

\looseness=-1
For the coordinated access, conditioned on $\mcI_t^{\rm C} \triangleq \{\text{pilot-activity history},\, \text{large-scale fading up to slot }t,\, \mcD_t\}$, the Kalman error covariance is deterministic and does not depend on the realized pilot observations. The latter are thus excluded from $\mcI_t^{\rm C}$. Unless stated otherwise, all expectations in this subsection are conditioned on $\mcI_t^{\rm C}$, and the slot index $t$ is omitted from intermediate expressions when no ambiguity arises. For user $k\in\mcG_r$, let $\widehat{\mathbf g}_{l}^{(k)}$ denote the Q-JCKF estimate at AP $l$ and $\mathbf e_l^{(k)} \triangleq \mathbf g_l^{(k)}-\widehat{\mathbf  g}_l^{(k)}$ the corresponding estimation error. Under the adopted linear-Gaussian AQNM model, the Q-JCKF is the conditional MMSE estimator; hence, $\E\!\big[ \widehat{\mathbf g}_l^{(k)} (\mathbf e_l^{(i)})^{\cH} \mid\mcI_t^{\rm C} \big] = \mathbf 0$ for $i,k\in\mcG_r$. The channel estimates and estimation errors are therefore conditionally independent, and the channel vectors and their estimates are zero-mean jointly proper Gaussian. Using the Kronecker structure $\mathbf P_{lr,t}^{+}
= \overline{\mathbf P}_{lr,t}^{+}\otimes\mathbf I_N$, let $p_{l,t}^{(i,k)}$ denote the $(i,k)$ entry of $\overline{\mathbf P}_{lr,t}^{+}$ for users $i,k\in\mcG_r$.  Then $\E\!\big[ \mathbf e_{l,t}^{(i)} (\mathbf e_{l,t}^{(k)})^{\cH} \mid\mcI_t^{\rm C} \big] = p_{l,t}^{(i,k)}\mathbf I_N$. Since the user channels are mutually independent before pilot processing, $\E\!\big[ \mathbf g_{l,t}^{(i)} (\mathbf g_{l,t}^{(k)})^{\cH} \mid\mcI_t^{\rm C} \big] = \delta_{ik}\beta_{l,t}^{(i)}\mathbf I_N$, and therefore $
\E\! \big[ \widehat{\mathbf g}_{l,t}^{(i)} (\widehat{\mathbf g}_{l,t}^{(k)})^{\cH} \mid\mcI_t^{\rm C} \big] = \big( \delta_{ik}\beta_{l,t}^{(i)} - p_{l,t}^{(i,k)} \big)\mathbf I_N$. Define the estimated-channel variance $\gamma_{l,t}^{(k)} \triangleq \beta_{l,t}^{(k)}-p_{l,t}^{(k,k)}$ and the estimate--channel cross-covariance coefficient $c_{l,t}^{(k,i)}
\triangleq \delta_{ik}\beta_{l,t}^{(i)}-p_{l,t}^{(i,k)}$, for $i,k\in\mcG_r$, so that $
\E\! \big[ \mathbf g_{l,t}^{(i)} (\widehat{\mathbf g}_{l,t}^{(k)})^{\cH} \mid\mcI_t^{\rm C} \big] = c_{l,t}^{(k,i)}\mathbf I_N$.
For users in different pilot groups, $c_{l,t}^{(k,i)}=0$ under the adopted groupwise Gaussian-AQNM model. Consequently, $ \E\! \big[ (\widehat{\mathbf g}_{l}^{(k)})^{\cH}
\mathbf g_l^{(i)} \mid\mcI_t^{\rm C} \big] = Nc_l^{(k,i)}$. Since $\widehat{\mathbf g}_{l}^{(k)}$ and $\mathbf g_l^{(i)}$ are jointly proper Gaussian,
\begin{equation}
\E\!\Big[ \big| (\widehat{\mathbf g}_{l}^{(k)})^{\cH} \mathbf g_l^{(i)} \big|^2
\mid\mcI_t^{\rm C} \Big] = N\gamma_l^{(k)}\beta_l^{(i)} + N^2|c_l^{(k,i)}|^2.
\label{eq:moments_coord}
\end{equation}
The $O(N)$ term is noncoherent, whereas the $O(N^2)$ term captures coherent estimate--channel correlation.

\looseness=-1
For $i\neq k$ in the same pilot group, $c_{l,t}^{(k,i)}=-p_{l,t}^{(i,k)}$. Thus, coherent same-pilot interference is governed directly by the off-diagonal Q-JCKF error covariances generated by the joint Riccati recursion in \eqref{eq:reduced_update}. These terms may remain nonzero even when an interfering user does not transmit a pilot in the current slot because the tracker retains information from earlier slots. Let $r(k)$ denote the pilot index assigned to user $k$, and define $\mathcal S_t^{p,(k)} \triangleq \big(\mcD_t\cap\mcG_{r(k)}\big)\setminus\{k\}$, the set of data-active users sharing the pilot of user $k$. With MRC, $\mathbf v_{l,t}^{(k)}=\widehat{\mathbf g}_{l,t}^{(k)}$, and \eqref{eq:effective_combined_channel} becomes $h_t^{(k,i)}= \sum_{l=1}^{L} \alpha_l (\widehat{\mathbf g}_{l,t}^{(k)})^{\cH} \mathbf g_{l,t}^{(i)}$. Under the independent-AP Gaussian-AQNM model, $\E\! \big[ h_t^{(k,i)}\mid\mcI_t^{\rm C} \big] = N\sum_{l=1}^{L}\alpha_l c_{l,t}^{(k,i)}$, while \eqref{eq:moments_coord} gives
\begin{equation}
\E\!\Big[ |h_t^{(k,i)}|^2 \mid\mcI_t^{\rm C}\Big] = N\sum_{l=1}^{L} \alpha_l^2\gamma_l^{(k)}\beta_l^{(i)} + N^2 \big| \sum_{l=1}^{L} \alpha_l c_l^{(k,i)} \big|^2. \label{eq:qjc_global_second_moment}
\end{equation}

\looseness=-1
For $i=k$, $c_l^{(k,k)}=\gamma_l^{(k)}$, yielding $\mu_t^{{\rm C},(k)} = N\sum_{l=1}^{L} \alpha_l\gamma_{l,t}^{(k)}$. Moreover, $\E[\|\widehat{\mathbf g}_{l,t}^{(k)}\|^2\mid\mcI_t^{\rm C}] = N\gamma_{l,t}^{(k)}$, so the mean effective receiver-noise power is
\begin{equation}
\bar\nu_t^{{\rm C},(k)} = N\sum_{l=1}^{L} \alpha_l \big[ 1+(1-\alpha_l)\rho_uB_{l,t} 
\big] \gamma_{l,t}^{(k)}. \label{eq:qjc_mean_noise}
\end{equation}

\begin{theorem} For a data-active user $k$, under independent Rayleigh fading, the Gaussian AQNM surrogate, known pilot activity, local MRC, and coherent CPU combining, the statistical UatF SINR conditioned on $\mcI_t^{\rm C}$ is
\begin{equation}
\mathrm{SINR}_t^{{\rm C},(k)} = \frac{ N\rho_u \big| \displaystyle\sum_{l=1}^{L} \alpha_l\gamma_{l,t}^{(k)}\big|^2 }{F_t^{(k)}+ N\rho_u \sum_{i\in\mathcal S_t^{p,(k)}} \big| \displaystyle\sum_{l=1}^{L} \alpha_l c_{l,t}^{(k,i)} \big|^2 }, \label{eq:sinr_c}
\end{equation}
 where $F_t^{(k)}\triangleq \sum_{l=1}^{L} \alpha_l\gamma_{l,t}^{(k)} (1+\rho_uB_{l,t})$. If all combining vectors vanish, the SINR is defined as zero. The corresponding UatF spectral-efficiency lower bound is
\begin{equation}
R_t^{{\rm C},(k)} = \eta_u \log_2 \big( 1+\mathrm{SINR}_t^{{\rm C},(k)}\big).
\label{eq:qjc_uatf_rate}
\end{equation}
\end{theorem}

\begin{IEEEproof}
Using \eqref{eq:qjc_global_second_moment},
\begin{align}
\sum_{i\in\mcD_t} \E\!\big[ |h_t^{(k,i)}|^2 \mid\mcI_t^{\rm C} \big] &=N\sum_{l=1}^{L} \alpha_l^2\gamma_l^{(k)}B_{l,t}+ N^2 \big| \sum_{l=1}^{L} \alpha_l\gamma_l^{(k)}\big|^2 \nonumber\\
&+ N^2 \sum_{i\in\mathcal S_t^{p,(k)}}\big| \sum_{l=1}^{L} \alpha_l c_l^{(k,i)} \big|^2 . \label{eq:qjc_total_second_moment}
\end{align}
After subtracting $\rho_u|\mu_t^{{\rm C},(k)}|^2$ and adding \eqref{eq:qjc_mean_noise}, the noncoherent part reduces to $N\sum_{l=1}^{L}
\alpha_l \gamma_l^{(k)} (1+\rho_uB_{l,t})$. Substitution into \eqref{eq:uatf_general} and division by $N$ give \eqref{eq:sinr_c}; the rate expression follows from the UatF definition above.
\end{IEEEproof}  

\looseness=-1
In \eqref{eq:sinr_c}, the first denominator term contains the noncoherent channel uncertainty, multiuser interference, thermal noise, and quantization distortion, whereas the second captures coherent interference from data-active same-pilot users. Since $c_{l,t}^{(k,i)}$ is determined by the posterior cross-covariances, this term depends on the complete pilot-activity history and channel dynamics, rather than only on pilots transmitted in the current slot.

\begin{corollary} Under the conditioning of Theorem~1, assume fixed $L$, transmit powers, ADC gains, large-scale fading, pilot-activity history, and per-antenna tracking statistics independent of $N$. Define $G_t^{(k)} \triangleq \sum_{l=1}^{L}\alpha_l\gamma_{l,t}^{(k)}$ and $I_t^{{\rm coh},(k)} \triangleq \sum_{i\in\mathcal {S}_t^{p,(k)}} \big| \sum_{l=1}^{L} \alpha_lc_{l,t}^{(k,i)} \big|^2$. For $G_t^{(k)}>0$, if $I_t^{{\rm coh},(k)}>0$, then
\begin{equation}
\lim_{N\rightarrow\infty} \mathrm{SINR}_t^{{\rm C},(k)} = \frac{ |G_t^{(k)}|^2 }{
I_t^{{\rm coh},(k)} }. \label{eq:qjc_sinr_ceiling}
\end{equation}
If $I_t^{{\rm coh},(k)}=0$, then
\begin{equation}
\mathrm{SINR}_t^{{\rm C},(k)} = N\rho_u \frac{ |G_t^{(k)}|^2 }{ F_t^{(k)} },
\label{eq:qjc_linear_scaling}
\end{equation}
and the SINR grows linearly with $N$.
\end{corollary}

\begin{IEEEproof}
For $I_t^{{\rm coh},(k)}>0$, divide \eqref{eq:sinr_c} by $N\rho_u$ and let $N\to\infty$, which gives \eqref{eq:qjc_sinr_ceiling}. Setting $I_t^{{\rm coh},(k)}=0$ in \eqref{eq:sinr_c} gives \eqref{eq:qjc_linear_scaling}. 
\end{IEEEproof}

\looseness=-1
A finite SINR ceiling requires $I_t^{{\rm coh},(k)}>0$; a nonzero local $c_{l,t}^{(k,i)}$ alone is insufficient because the weighted AP contributions may cancel after coherent CPU combining. Hence, persistent coherent interference cannot be removed by increasing $N$, whereas the SINR grows linearly with $N$ when this term vanishes. This also explains why the off-diagonal tracking-error covariances must be retained in the rate analysis. The expected special cases follow directly: $\alpha_l=1$ gives ideal ADCs; perfect tracking $\overline{\mathbf P}_{lr,t}^{+}=\mathbf0$ gives $\gamma_l^{(k)} =\beta_l^{(k)}$ and $c_l^{(k,i)}=0$ for $i\neq k$; and without pilot reuse, $\mathcal{S}_t^{p,(k)}=\varnothing$, so the coherent-interference term vanishes.

\subsection{Q-DPDAF Mixture-Aware UatF Analysis}

\looseness=-1
Unlike the coordinated Q-JCKF, Q-DPDAF retains uncertainty over the pilot-activity hypotheses. For current-slot rate evaluation, it uses the Gaussian-mixture posterior in \eqref{eq:mixture}. We condition on the pilot information $\mcY_t$, the data-active set $\mcD_t$, and the known large-scale fading, with the latter two omitted from the conditioning notation. Posterior expectations are denoted by $\E_{\rm D}[\cdot]$. For user $k$ assigned to pilot group $r(k)$, the Q-DPDAF channel estimate at AP $l$ is
\begin{equation}
\widehat{\mathbf g}_{l,t}^{{\rm D},(k)} = \E_{\rm D}\!\big[ \mathbf g_{l,t}^{(k)} \mid\mcY_t \big] = \sum_{\mathbf q\in\mcQ_{r(k)}} \omega_{r(k),t}(\mathbf q)
\mathbf m_{l,t}^{(k)}(\mathbf q), \label{eq:dpdaf_channel_estimate_rate}
\end{equation}
where $\mathbf m_{l,t}^{(k)}(\mathbf q)$ is the user-$k$ component of $\mathbf m_{l,r(k),t}^{+}(\mathbf q)$. With MRC, $\mathbf v_{l,t}^{(k)} = \widehat{\mathbf g}_{l,t}^{{\rm D},(k)}$. Conditioned on $\mcY_t$, these combining vectors are deterministic; hence, \eqref{eq:effective_combined_channel} applies directly.

\looseness=-1
For a data-active user $i\in\mcD_t$ assigned to pilot group $r(i)$ and $\mathbf q\in\mcQ_{r(i)}$, let $\mathbf P_{l,t}^{(i,i)}(\mathbf q)$ denote the user-$i$ diagonal block of $\mathbf P_{l,r(i),t}^{+}(\mathbf q)$. The hypothesis-conditioned channel at AP $l$ is then $\mathbf g_{l,t}^{(i)} = \mathbf m_{l,t}^{(i)}(\mathbf q) + \mathbf e_{l,t}^{(i)}(\mathbf q)$, where $\mathbf e_{l,t}^{(i)}(\mathbf q) \sim \CN\!\big( \mathbf 0, \mathbf P_{l,t}^{(i,i)}(\mathbf q) \big)$. Under the factorized Gaussian-mixture model, these errors are zero-mean and independent across APs conditioned on
$(\mathbf q,\mcY_t)$. Define $a_{l,t}^{(k,i)}(\mathbf q) \triangleq (\mathbf v_{l,t}^{(k)})^{\cH} \mathbf m_{l,t}^{(i)}(\mathbf q)$, $u_{l,t}^{(k,i)}(\mathbf q)  \triangleq (\mathbf v_{l,t}^{(k)})^{\cH} \mathbf P_{l,t}^{(i,i)}(\mathbf q) \mathbf v_{l,t}^{(k)} \geq0$, and $ A_t^{(k,i)}(\mathbf q) \triangleq \sum_{l=1}^{L} \alpha_l a_{l,t}^{(k,i)}(\mathbf q)$. Hence,
$\E_{\rm D}\!\big[ |h_t^{(k,i)}|^2 \mid\mathbf q,\mcY_t \big] = |A_t^{(k,i)}(\mathbf q)|^2 + \sum_{l=1}^{L} \alpha_l^2 u_{l,t}^{(k,i)}(\mathbf q).$

\looseness=-1
Define $\overline A_t^{(k,i)} \triangleq \sum_{\mathbf q\in\mcQ_{r(i)}} \omega_{r(i),t}(\mathbf q) A_t^{(k,i)}(\mathbf q)$. Marginalizing the shared activity hypothesis then gives
\begin{align}
\Upsilon_t^{(k,i)} &\triangleq \E_{\rm D}\!\big[ |h_t^{(k,i)}|^2 \mid\mcY_t \big]
\nonumber\\ 
&= \sum_{\mathbf q\in\mcQ_{r(i)}} \omega_{r(i),t}(\mathbf q) \Big[
|A_t^{(k,i)}(\mathbf q)|^2 + \sum_{l=1}^{L} \alpha_l^2 u_{l,t}^{(k,i)}(\mathbf q)
\Big]. \label{eq:Upsilon}
\end{align}
Because the coherent AP sum is formed for each shared activity hypothesis before marginalization, averaging $|A_t^{(k,i)}(\mathbf q)|^2$ retains the cross-AP association induced by the common group hypothesis. Equivalently,
$\Upsilon_t^{(k,i)} = |\overline A_t^{(k,i)}|^2 + \sum_{\mathbf q\in\mcQ_{r(i)}} \omega_{r(i),t}(\mathbf q) \big| A_t^{(k,i)}(\mathbf q) - \overline A_t^{(k,i)} \big|^2 \; + \sum_{\mathbf q\in\mcQ_{r(i)}} \omega_{r(i),t}(\mathbf q) \sum_{l=1}^{L} \alpha_l^2 u_{l,t}^{(k,i)}(\mathbf q),$
where the last two terms represent uncertainty across activity hypotheses and within each hypothesis, respectively.

\looseness=-1
Using \eqref{eq:dpdaf_channel_estimate_rate}, MRC gives $\overline A_t^{(k,k)} = \sum_{l=1}^{L} \alpha_l\|\mathbf v_{l,t}^{(k)}\|^2 \ge0$. Since the combining vectors are fixed conditioned on $\mcY_t$, the data-phase AQNM gives $B_{l,t} = \sum_{i\in\mcD_t} \beta_{l,t}^{(i)}$ and 
$\nu_t^{(k)} = \sum_{l=1}^{L} \alpha_l \left[ 1+(1-\alpha_l)\rho_u B_{l,t} \right] \|\mathbf v_{l,t}^{(k)}\|^2.$

\begin{theorem} \label{thm:dpdaf_uatf}
Under independent Gaussian signaling, local MRC, the factorized Gaussian-mixture posterior in \eqref{eq:mixture}, and the adopted Gaussian-AQNM data model, the conditional UatF SINR of a data-active user $k\in\mcD_t$ is
\begin{equation}
\mathrm{SINR}_{t}^{{\rm D}\mid\mcY,(k)} = \frac{ \rho_u |\overline A_t^{(k,k)}|^2 }{\rho_u \displaystyle\sum_{i\in\mcD_t} \Upsilon_t^{(k,i)} - \rho_u |\overline A_t^{(k,k)}|^2  + \nu_t^{(k)} }. \label{eq:sinr_d}
\end{equation}
If all combining vectors for user $k$ vanish, the SINR is defined as zero. The corresponding conditional UatF spectral-efficiency lower bound is
\begin{equation}
R_t^{{\rm D}\mid\mcY,(k)} = \eta_u \log_2 \big( 1+ \mathrm{SINR}_{t}^{{\rm D}\mid\mcY,(k)} \big). \label{eq:dpdaf_cond_rate}
\end{equation}
\end{theorem}

\begin{IEEEproof}
Conditioned on $\mcY_t$, the desired effective-channel gain has mean $\overline A_t^{(k,k)}$ and variance $\Upsilon_t^{(k,k)} -|\overline A_t^{(k,k)}|^2$. Independent zero-mean symbols give multiuser-interference power $\rho_u\sum_{i\in\mcD_t,\;i\neq k} \Upsilon_t^{(k,i)}$. Adding these terms to $\nu_t^{(k)}$ gives the denominator of \eqref{eq:sinr_d}. The worst-case uncorrelated-noise argument gives \eqref{eq:sinr_d}, and the payload factor $\eta_u$ gives \eqref{eq:dpdaf_cond_rate}.
\end{IEEEproof}

\looseness=-1
Theorem~\ref{thm:dpdaf_uatf} retains pilot-activity uncertainty through the posterior weights and requires no hard activity decision before rate evaluation. It is a model-based UatF bound under the Gaussian-mixture/AQNM approximation of Section~III-B; no exact achievable-rate claim is made for the original nonlinear quantized channel. For comparison with the statistical Q-JCKF bound, define a statistical Q-DPDAF UatF bound by averaging over the random pilot-activity and pilot-observation histories, while conditioning on the prescribed large-scale-fading trajectory and $\mcD_t$. Let $\E_{\rm hist}[\cdot]$ denote this averaging. Then $\mu_t^{{\rm D},(k)} \triangleq \E_{\rm hist}\!\big[ \overline A_t^{(k,k)} \big]$, $\Psi_t^{{\rm D},(k,i)} \triangleq \E_{\rm hist}\!\big[ \Upsilon_t^{(k,i)} \big]$, and $\overline\nu_t^{{\rm D},(k)} \triangleq \E_{\rm hist}\!\big[ \nu_t^{(k)} \big]$. The resulting statistical UatF SINR is
\begin{equation}
\mathrm{SINR}_t^{{\rm D},(k)} = \frac{ \rho_u|\mu_t^{{\rm D},(k)}|^2 }{ \rho_u \displaystyle\sum_{i\in\mcD_t} \Psi_t^{{\rm D},(k,i)} - \rho_u|\mu_t^{{\rm D},(k)}|^2 + \overline\nu_t^{{\rm D},(k)} }. \label{eq:det_d}
\end{equation}
with $R_t^{{\rm D},(k)} = \eta_u \log_2 \big( 1+\mathrm{SINR}_t^{{\rm D},(k)}
\big)$. Unlike the coordinated expression, these moments generally require numerical averaging because the posterior weights depend nonlinearly on the random pilot observations through the AP likelihoods.

\looseness=-1
In general, $\E_{\rm hist}\!\big[ R_t^{{\rm D}\mid\mcY,(k)} \big] \neq R_t^{{\rm D},(k)}$: the former averages rates formed from pilot-conditioned moments, whereas the latter forms one SINR from statistically averaged moments. For the fixed MRC vectors above, a joint Gaussian approximation preserving the full posterior mean and covariance reproduces the same UatF first and second moments. In contrast, independent AP-local moment matching discards the cross-AP association term quantified in Section~IV-C.

\subsection{Cross-AP Association Analysis}

\looseness=-1
The mixture-aware SINR in \eqref{eq:sinr_d} preserves the group-activity hypothesis shared across the APs. Although the AP states are conditionally independent for a fixed hypothesis under the adopted factorized model, marginalizing this hypothesis induces cross-AP dependence. We characterize this association covariance, its contribution to the UatF second moment, and its dependence on posterior ambiguity. For a data-active user $i\in\mcD_t$ assigned to pilot group $r(i)$, define $\omega^{(i)}(\mathbf q) \triangleq \omega_{r(i),t}(\mathbf q)$. All quantities below are conditioned on $\mcY_t$, $\mcD_t$, and the known large-scale fading, with the latter two omitted from the notation. The slot index is suppressed when clear. Under hypothesis $\mathbf q$, let $\mathbf m_l^{(i)}(\mathbf q)$ and $\mathbf P_l^{(i,i)}(\mathbf q)$ denote the posterior mean and covariance of $\mathbf g_l^{(i)}$. Define $\overline{\mathbf m}_l^{(i)} \triangleq \sum_{\mathbf q\in\mcQ_{r(i)}} \omega^{(i)}(\mathbf q) \mathbf m_l^{(i)}(\mathbf q)$ and $\mathbf \Delta_l^{(i)}(\mathbf q) \triangleq \mathbf m_l^{(i)}(\mathbf q) - \overline{\mathbf m}_l^{(i)}$. Then $\sum_{\mathbf q}\omega^{(i)}(\mathbf q) \mathbf\Delta_l^{(i)}(\mathbf q)=\mathbf0$ and, by the law of total covariance,
$\overline{\mathbf P}_l^{(i,i)} ={} \sum_{\mathbf q\in\mcQ_{r(i)}} \omega^{(i)}(\mathbf q) \mathbf P_l^{(i,i)}(\mathbf q) \; + \sum_{\mathbf q\in\mcQ_{r(i)}} \omega^{(i)}(\mathbf q) \mathbf\Delta_l^{(i)} (\mathbf q) \big( \mathbf\Delta_l^{(i)}(\mathbf q) \big)^{\cH}.$
The two terms present, respectively, within-hypothesis estimation uncertainty and uncertainty across activity hypotheses. Equivalently, $\overline{\mathbf m}_l^{(i)}$ and
$\overline{\mathbf P}_l^{(i,i)}$ are the user-$i$ mean block and diagonal covariance block of $\mathbf m_{l,r(i),t}^{\rm MM}$ and $\mathbf P_{l,r(i),t}^{\rm MM}$, respectively.

\looseness=-1
For two distinct APs $l$ and $m$, define $\mathbf C_{lm}^{{\rm assoc},(i)} \triangleq \E_{\rm D}\!\big[ (\mathbf g_l^{(i)}-\overline{\mathbf m}_l^{(i)}) (\mathbf g_m^{(i)}-\overline{\mathbf m}_m^{(i)})^{\cH} \mid\mcY_t \big]$. Under the factorized posterior, $\operatorname{Cov}_{\rm D} (\mathbf g_l^{(i)},\mathbf g_m^{(i)} \mid\mathbf q,\mcY_t)=\mathbf0$ for $l\neq m$. Thus, the law of total covariance gives
\begin{equation}
\mathbf C_{lm}^{{\rm assoc},(i)} = \sum_{\mathbf q\in\mcQ_{r(i)}} \omega^{(i)}(\mathbf q) \mathbf\Delta_l^{(i)}(\mathbf q) \big(\mathbf\Delta_m^{(i)}(\mathbf q)\big)^{\cH}, \label{eq:Cassoc}
\end{equation}
where \eqref{eq:Cassoc} captures dependence induced solely by the shared activity hypothesis under the adopted factorized posterior; it is distinct from physical spatial channel correlation or correlated receiver noise. As Q-DPDAF propagates only AP-local marginals, \eqref{eq:Cassoc} represents current-slot association rather than the full
cross-AP dependence of an exact joint Bayesian filter.

\looseness=-1
We quantify its contribution to the UatF second moment. For fixed MRC vectors, $h^{(k,i)} = \sum_{l=1}^{L} \alpha_l (\mathbf v_l^{(k)})^{\cH} \mathbf g_l^{(i)}$ and $\overline A^{(k,i)} = \sum_{l=1}^{L} \alpha_l (\mathbf v_l^{(k)})^{\cH} \overline{\mathbf m}_l^{(i)}$. Using \eqref{eq:Cassoc} and $\mathbf C_{ml}^{{\rm assoc},(i)}= (\mathbf C_{lm}^{{\rm assoc},(i)})^{\cH}$ gives 
\begin{equation}
\begin{aligned}
\Upsilon^{(k,i)} ={}& |\overline A^{(k,i)}|^2 + \sum_{l=1}^{L} \alpha_l^2 (\mathbf v_l^{(k)})^{\cH} \overline{\mathbf P}_l^{(i,i)} \mathbf v_l^{(k)} \\
&+ 2\Re\!\Big\{ \sum_{l<m} \alpha_l\alpha_m (\mathbf v_l^{(k)})^{\cH} \mathbf C_{lm}^{{\rm assoc},(i)} \mathbf v_m^{(k)} \Big\}. \end{aligned} \label{eq:assoc_decomp}
\end{equation}
The three terms are the squared posterior mean, AP-local posterior uncertainty, and cross-AP association contribution, respectively.

\looseness=-1
To quantify the approximation error caused by discarding this association, consider, for $r=r(i)$, the product of the AP-local moment-matched Gaussian marginals defined in Section~III-B,
$p_{\rm LM} \big( \mathbf x_{1r,t},\ldots,\mathbf x_{Lr,t} \mid\mcY_t \big) = \prod_{l=1}^{L} \CN\!\big( \mathbf x_{lr,t}; \mathbf m_{lr,t}^{\rm MM}, \mathbf P_{lr,t}^{\rm MM} \big).$
This approximation preserves the marginal first and second moments at each AP but discards the cross-AP association covariance. Its effective-channel second moment is therefore
$\Upsilon^{{\rm LM},(k,i)} = |\overline A^{(k,i)}|^2 + \sum_{l=1}^{L} \alpha_l^2 (\mathbf v_l^{(k)})^{\cH} \overline{\mathbf P}_l^{(i,i)} \mathbf v_l^{(k)}.$
Thus,
\begin{equation}
\begin{aligned}
J^{{\rm assoc},(k,i)} &\triangleq \Upsilon^{(k,i)} - \Upsilon^{{\rm LM},(k,i)} \\
&= 2\Re\!\Big\{ \sum_{l<m} \alpha_l\alpha_m (\mathbf v_l^{(k)})^{\cH} \mathbf C_{lm}^{{\rm assoc},(i)} \mathbf v_m^{(k)} \Big\}. \end{aligned}
\label{eq:assoc_correction}
\end{equation}

\looseness=-1
The correction $J^{{\rm assoc},(k,i)}$ can be positive or negative because the AP contributions may combine constructively or destructively. Thus, independent AP-local moment matching gives neither a universal upper nor a universal lower bound on the mixture-aware second moment or UatF rate. In contrast, any joint approximation preserving the complete posterior mean and covariance yields the same fixed-combiner UatF first and second moments, although the underlying distributions need not be identical. The association covariance can be related directly to posterior ambiguity.

\begin{lemma} For user $i$, define $\eps^{(i)} \triangleq 1- \max_{\mathbf q\in\mcQ_{r(i)}} \omega^{(i)}(\mathbf q)$ and $D_l^{(i)} \triangleq \max_{\mathbf q,\mathbf q'\in\mcQ_{r(i)}}  \big\| \mathbf m_l^{(i)}(\mathbf q) - \mathbf m_l^{(i)}(\mathbf q') \big\|_2$. For $l\neq m$,
\begin{align}
\big\| \mathbf C_{lm}^{{\rm assoc},(i)} \big\|_{\rm F} &\leq \big( \eps^{(i)} -
\frac{(\eps^{(i)})^2}{2} \big) D_l^{(i)}D_m^{(i)}, \nonumber\\ 
\frac{ \big\| \mathbf C_{lm}^{{\rm assoc},(i)} \big\|_{\rm F} }{ D_l^{(i)}D_m^{(i)}
} &\leq \eps^{(i)} - \frac{(\eps^{(i)})^2}{2}, \qquad D_l^{(i)}D_m^{(i)}>0.
\label{eq:eps_bound}
\end{align}
If $D_l^{(i)}D_m^{(i)}=0$, then $\mathbf C_{lm}^{{\rm assoc},(i)}=\mathbf0$. For uniformly bounded diameters, $\| \mathbf C_{lm}^{{\rm assoc},(i)} \|_{\rm F} =
\mathcal O(\eps^{(i)})$ as $\eps^{(i)}\rightarrow0$.
\end{lemma}

\begin{IEEEproof}
Let $\omega_{\mathbf q}\triangleq\omega^{(i)}(\mathbf q)$. The pairwise covariance identity gives
\begin{align}
\mathbf C_{lm}^{{\rm assoc},(i)} = \frac{1}{2} \sum_{\mathbf q,\mathbf q'} \omega_{\mathbf q}\omega_{\mathbf q'} & \Big[ \mathbf m_l^{(i)}(\mathbf q) -
\mathbf m_l^{(i)}(\mathbf q') \Big] \nonumber\\[-1mm] &\times \Big[ \mathbf m_m^{(i)}(\mathbf q) - \mathbf m_m^{(i)}(\mathbf q') \Big]^{\cH}. \label{eq:assoc_pairwise_identity}
\end{align}
Therefore, $\big\| \mathbf C_{lm}^{{\rm assoc},(i)} \big\|_{\rm F} \leq \frac{D_l^{(i)}D_m^{(i)}}{2} \big( 1-\sum_{\mathbf q}\omega_{\mathbf q}^{2} \big)$.
Since $\max_{\mathbf q}\omega_{\mathbf q}=1-\eps^{(i)}$, we have $\sum_{\mathbf q}\omega_{\mathbf q}^{2} \geq (1-\eps^{(i)})^2$. Substitution gives the first inequality in \eqref{eq:eps_bound}; division by $D_l^{(i)}D_m^{(i)}$ gives the second when this product is nonzero. If $D_l^{(i)}D_m^{(i)}=0$, the pairwise identity implies
$\mathbf C_{lm}^{{\rm assoc},(i)}=\mathbf0$.
\end{IEEEproof}

\looseness=-1
Thus, the association magnitude is controlled jointly by posterior ambiguity and hypothesis-dependent mean separation. A diffuse posterior need not produce large association when the conditional means remain close. The same result bounds the omitted contribution in \eqref{eq:assoc_correction}. Using Cauchy--Schwarz and $\|\mathbf C\|_2\leq\|\mathbf C\|_{\rm F}$,
\begin{equation}
\begin{aligned}
\big| J^{{\rm assoc},(k,i)} \big| \leq{}& 2 \Big( \eps^{(i)} - \frac{(\eps^{(i)})^2}{2} \Big) \\
&\times \sum_{l<m} \alpha_l\alpha_m \|\mathbf v_l^{(k)}\|_2 \|\mathbf v_m^{(k)}\|_2 D_l^{(i)}D_m^{(i)}. \end{aligned} \label{eq:assoc_scalar_bound}
\end{equation}
Hence, $J^{{\rm assoc},(k,i)} = \mathcal O(\eps^{(i)})$ for bounded combining vectors and conditional-mean diameters.

\looseness=-1
To quantify the rate effect, define $\mathcal J^{(k)} \triangleq \sum_{i\in\mcD_t} J^{{\rm assoc},(k,i)}$ and 
\[ \begin{aligned}
\mathcal V^{{\rm mix},(k)} &\triangleq \rho_u\sum_{i\in\mcD_t}\Upsilon^{(k,i)} -\rho_u|\overline A^{(k,k)}|^2+\nu^{(k)},\\
\mathcal V^{{\rm LM},(k)} &\triangleq \rho_u\sum_{i\in\mcD_t}\Upsilon^{{\rm LM},(k,i)} -\rho_u|\overline A^{(k,k)}|^2+\nu^{(k)}.
\end{aligned} \]
Then $\mathcal V^{{\rm mix},(k)} = \mathcal V^{{\rm LM},(k)} + \rho_u\mathcal J^{(k)}$.

\looseness=-1
For $x\in\{{\rm mix},{\rm LM}\}$, let $R^{x,(k)} \triangleq \eta_u \log_2 \big( 1+
\frac{S^{(k)}}{\mathcal V^{x,(k)}} \big)$ and $S^{(k)} \triangleq \rho_u|\overline A^{(k,k)}|^2$. If both disturbance powers are positive, the mean-value theorem
gives
\begin{equation}
\Big| R^{{\rm mix},(k)} - R^{{\rm LM},(k)} \Big| \leq \frac{ \eta_u S^{(k)} \rho_u| \mathcal J^{(k)}| }{ \ln(2)\, \mathcal V_{\min} \big( \mathcal V_{\min}+S^{(k)} \big)
}, \label{eq:assoc_rate_gap_bound}
\end{equation}
where $\mathcal V_{\min} \triangleq \min\{ \mathcal V^{{\rm mix},(k)}, \mathcal V^{{\rm LM},(k)} \}>0$. For a fixed finite active-user set and uniformly bounded quantities, the rate gap therefore vanishes as $\eps^{(i)}\to0$ for all $i\in\mcD_t$. Neither the sign of $\mathcal J^{(k)}$ nor a universal ordering of the two rates follows; each user has its own pilot group and ambiguity $\eps^{(i)}$, so no single ambiguity parameter characterizes the complete rate correction.

\looseness=-1
The $\mathcal O(\epsilon^{(i)})$ result concerns only the current-slot cross-AP dependence induced by the shared activity hypothesis; it does not characterize the total Q-DPDAF tracking, AQNM, or outside-group mean-field errors. For uniformly bounded conditional-mean diameters, $\omega^{(i)}(\mathbf q^\star)\to1$ implies $\mathbf C_{lm}^{{\rm assoc},(i)}\to\mathbf0$, so the mixture-aware and AP-local moment-matched second moments coincide for that user. If this concentration holds for all $i\in\mathcal D_t$, the corresponding UatF rate gap also vanishes. Posterior concentration alone does not make Q-DPDAF equivalent to Q-JCKF. Exact agreement of the conditional updates additionally requires $\mathbf q^\star=\mathbf q_{r,t}$ and identical predicted means, covariances, and effective measurement-noise covariances. With finite-resolution ADCs, the latter also requires $\widehat{\sigma}_{lr,t}^{2}(\mathbf q^\star)=\sigma_{l,t}^{2}(\mathbf q_t)$; the outside-group mean-field mismatch disappears as $\alpha_l\to1$. Different past activity information may still produce different predicted states.

\subsection{Long-Term Statistical UatF Estimation}

\looseness=-1
For the long-term statistical UatF evaluation, replicas share the network layout and prescribed slow-time large-scale-fading trajectory but use independent pilot-activity, temporally correlated small-scale-fading, and receiver-noise realizations. The data-active set is fixed over the retained slots and replicas, consistent with Section IV. The required effective-channel moments and effective receiver-noise power are pooled over the retained post-burn-in slots and replicas and inserted into \eqref{eq:uatf_general} before forming the SINR and rate. The outer Monte Carlo average retains the realized binary pilot-activity sequences, whereas the Q-DPDAF recursion itself still uses the outside-group mean-field variance in \eqref{eq:dpdaf_meanfield_variance}. Thus, averaging is performed at the moment level rather than the rate level. To estimate the squared coherent desired gain without self-product bias, let
$a\in\{{\rm C},{\rm D}\}$ denote the Q-JCKF or Q-DPDAF receiver, respectively. We use the cross-replica estimator
\begin{equation}
\widehat{G}^{a,(k)} = \frac{ \big| \sum_{n=1}^{N_{\rm rep}} \widehat{\mu}_{n}^{a,(k)} \big|^2 - \sum_{n=1}^{N_{\rm rep}} \big| \widehat{\mu}_{n}^{a,(k)} \big|^2
}{ N_{\rm rep}(N_{\rm rep}-1) }, \label{eq:lt_cross_replica_gain}
\end{equation}
where $\widehat{\mu}_{n}^{a,(k)}$ is the per-replica estimate of the desired effective-channel mean obtained from the retained slots. Conditioned on the common layout and slow-time trajectory, replica independence gives $\E[\widehat{G}^{a,(k)}] = |\E[\widehat{\mu}_{n}^{a,(k)}]|^2$ before clipping. Negative finite-sample values of $\widehat{G}^{a,(k)}$ are clipped to zero, and a small positive floor is applied only to the estimated disturbance power for numerical stability.

\section{Complexity Analysis}

For pilot group $r$, the AP-local state and despread-observation dimensions are $NK_r$ and $N$, respectively. We characterize the leading-order computational scaling of both trackers and the hypothesis-induced memory and fronthaul costs of Q-DPDAF.

\subsection{Q-JCKF}

\looseness=-1
The Kronecker covariance structure established in Section~III-A reduces the $NK_r\times NK_r$ covariance recursion to the $K_r\times K_r$ user domain. For one AP and one pilot group, the reduced covariance update requires $\mathcal O(K_r^2)$ operations, while the state-mean update requires $\mathcal O(NK_r)$. Hence, the network-wide complexity is $C^{\rm Q-JCKF} = \mathcal O\!\big[ L\sum_{r=1}^{\tau_p} \big(K_r^2+NK_r\big) \big]$. If the Kronecker reduction is not exploited, the dense covariance correction scales as $\mathcal{O}(N^{3}K_r^{2})$ per AP and pilot group. The corresponding balanced-group scalings are given in Table I.

\subsection{Q-DPDAF}

\looseness=-1
Without hypothesis pruning, Q-DPDAF evaluates all $2^{K_r}$ in-group activity hypotheses. After Gaussian-mixture moment matching, the between-hypothesis covariance term generally destroys the Q-JCKF Kronecker structure; thus, Q-DPDAF propagates a full $NK_r \times NK_r$ AP-local covariance. For one hypothesis, one AP, and one pilot group, the dominant full-covariance correction scales as $\mathcal O(N^3K_r^2)$. Therefore, $C^{\rm Q-DPDAF} = \mathcal O\! \big[ LN^3 \sum_{r=1}^{\tau_p} 2^{K_r}K_r^2\big]$. The corresponding balanced-group scaling is summarized in Table I. CPU fusion of the AP-local log-likelihoods requires $\mathcal O\! \big( L\sum_{r=1}^{\tau_p}2^{K_r} \big)$ scalar operations and does not alter the leading Q-DPDAF scaling. Simultaneous storage of all hypothesis-conditioned covariances requires $\mathcal O\! \big[ LN^2 \sum_{r=1}^{\tau_p} 2^{K_r}K_r^2 \big]$ memory. The AP-to-CPU likelihood exchange requires $\mathcal O\!\big( L\sum_{r=1}^{\tau_p}2^{K_r} \big)$ real scalars per slot; returning all posterior weights separately to the APs has the same order. Under the factorized AP-local Q-DPDAF model of Section III-B, the hypothesis-conditioned AP updates are separable and can be processed in parallel. With parallel AP updates and serial CPU fusion, the idealized per-group computational critical path scales as $\mathcal O\! \big( 2^{K_r}N^3K_r^2 + L2^{K_r}\big)$. This idealized parallel scaling differs from the single-process MATLAB runtime reported in Section VI (Fig. 6), where the AP-local updates are executed serially.

\begin{table}[t]
\centering
\caption{Leading-Order Computational Complexity}
\label{tab:complexity}
\footnotesize
\begin{tabular}{@{}lcc@{}}
\toprule
Algorithm & Per AP/group & Balanced-network total \\
\midrule 
Q-JCKF & $\mathcal O(K_r^2+NK_r)$ & $\mathcal O\!\big(L(K^2/\tau_p+NK)\big)$ \\
Q-JCKF (unstructured)  & $\mathcal O(N^3K_r^2)$ & $\mathcal O(LN^3K^2/\tau_p)$ \\
Q-DPDAF & $\mathcal O(2^{K_r}N^3K_r^2)$ & $\mathcal O\!\big( LN^3K^2 2^{K/\tau_p}/\tau_p \big)$ \\
\bottomrule
\end{tabular}
\end{table}

\looseness=-1
Under the ideal-ADC co-located reduction of Section III-C, $\alpha_{l}$ approaches $1$, $L=1$, $N=M$, and $K_r=K$, with otherwise matched model parameters. The corresponding full-matrix JC-KF and PDAF complexities scale as $\mathcal O(M^3 K^2)$ and $\mathcal O(2^K M^3K^2)$, respectively. The predicted dependence on $K_r$ is consistent with the single-process MATLAB trends in Section VI (Fig.~\ref{fig:fig6}) and with \cite{SFR21}.

\section{Numerical Results}

\begin{table}[!t]
\centering
\caption{Simulation Parameters Used in the Numerical Results}
\label{tab:simparams}
\footnotesize
\begin{tabular}{@{}ll@{}}
\toprule
Parameter & Value(s) \\
\midrule
Users $K$ & 30 \\
APs $L$ & 25 \\
Antennas/AP $N$ & 4 \\
Total antennas & 100 \\
Pilot length $\tau_p$ & 5 \\
Group size $K_r$ & 6 \\
Slot length $\tau_{\rm sl}$ & 100 \\
Payload length $\tau_u$ & 95 \\
Pilot activity $\lambda_k$ & 0.5 \\
ADC resolution & 2 bits \\
Mobility speeds & $0,5,15,30$ m/s \\
Pilot/data transmit power & $100~\mathrm{mW}$ each \\
Path-loss exponent & 3.5 in geometry studies \\
Shadowing std. & 8 dB when enabled \\
\bottomrule
\end{tabular}
\end{table}

\looseness=-1
Unless stated otherwise, the simulations use the parameters in Table~II, with $\rho_t^{(k)}=\rho=0.95$ when a fixed temporal correlation is imposed. The uplink employs MRC with coherent CPU combining, and all reported spectral efficiencies are UatF quantities under the adopted Gaussian-AQNM surrogate. Users are data-active during rate evaluation unless stated otherwise. For a 180-kHz bandwidth, a 9-dB receiver noise figure, $T=290$ K, and 100-mW transmit power, receiver-noise normalization gives $\rho_p=\rho_u\simeq1.75\times10^{13}$. These quantities are normalized transmit powers rather than received SNRs; AP-specific received powers are further scaled by the corresponding large-scale fading coefficients. Long-term UatF moments are evaluated as in Section~IV-D, while deployment uncertainty is assessed over independent network geometries.

\begin{figure}[!t]
\centering
\subfloat[Rate validation.]{\includegraphics[width=0.47\columnwidth]{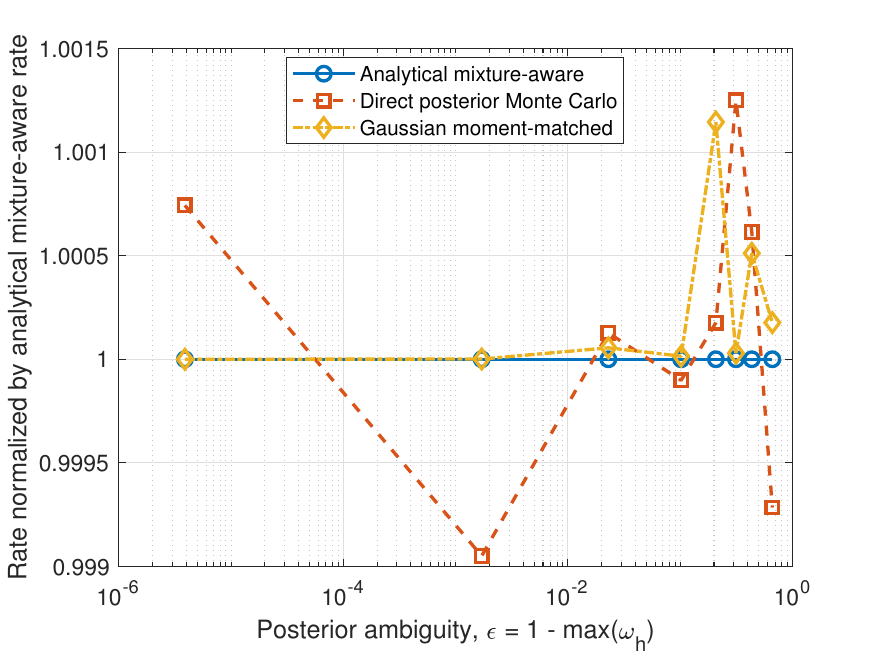}\label{fig:fig1a}}%
\hfill
\subfloat[Cross-AP association.]{\includegraphics[width=0.47\columnwidth]{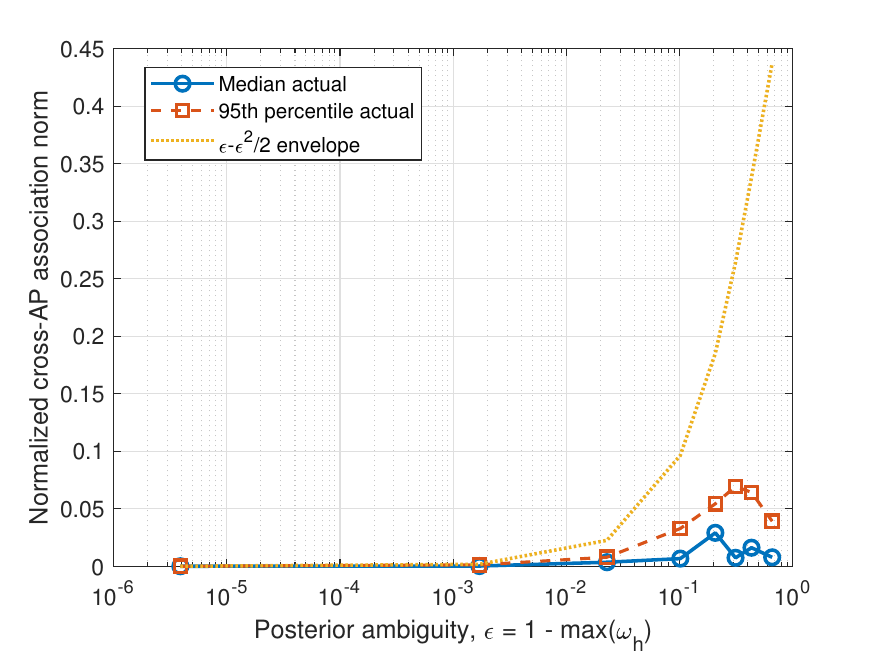}\label{fig:fig1b}}%
\caption{Numerical validation of the mixture-aware Q-DPDAF UatF analysis versus posterior ambiguity $\epsilon$.}
\label{fig:fig1}
\end{figure}

\looseness=-1
Fig.~\ref{fig:fig1} validates the mixture-aware Q-DPDAF UatF formulation of Section~IV-B, the cross-AP association analysis of Section~IV-C, and the AP-local moment-matched approximation. To isolate these effects from the outside-group mean-field approximation, we set $K=K_r=6$, $\tau_p=1$, $L=8$, $N=4$, and use 2-bit ADCs, yielding $2^{K_r}=64$ activity hypotheses. Posterior ambiguity $\epsilon=1-\max_q\omega(q)$ is varied through $\rho_p$ while $\rho_u$ remains fixed. Each posterior realization draws a common activity hypothesis according to $\omega(q)$ and conditionally independent AP states from the corresponding Gaussian posteriors, thus retaining the dependence induced by the shared hypothesis after marginalization. In Fig.~\ref{fig:fig1a}, direct posterior Monte Carlo closely matches the analytical mixture-aware rate. Over 80 posterior snapshots with 15,000 draws each, the median, 95th-percentile, and maximum relative rate errors are approximately $0.133\%$, $0.502\%$, and $1.00\%$, respectively. The AP-local moment-matched approximation is also highly accurate, with median and 95th-percentile relative errors of approximately $3.0\times10^{-6}$ and $5.9\times10^{-4}$. Fig.~\ref{fig:fig1b} independently verifies the association analysis: the analytical decomposition agrees with direct numerical evaluation to within $1.3\times10^{-15}$, and the normalized association magnitude remains below the theoretical envelope $\epsilon-\epsilon^2/2$. Across 2160 posterior samples, the maximum actual-to-bound ratio remains below unity. The omitted association magnitude and the moment-matched rate error exhibit a Spearman correlation of approximately $0.996$, while the rate error remains below $1\%$ in $98.9\%$ of samples and below $5\%$ in all samples. These results support both the association bound and the accuracy of the AP-local approximation over the tested ambiguity range.

\begin{figure}[!t]
\centering
\subfloat[Analytical UatF validation.]{\includegraphics[width=0.47\columnwidth]{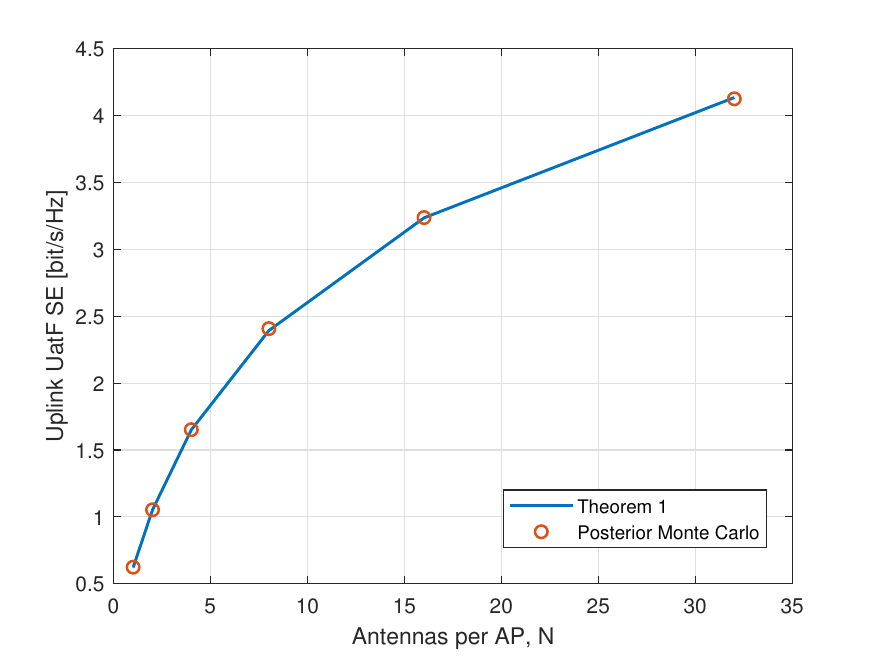}\label{fig:fig2a}}%
\hfill
\subfloat[Large-$N$ pilot-reuse behavior.]{\includegraphics[width=0.47\columnwidth]{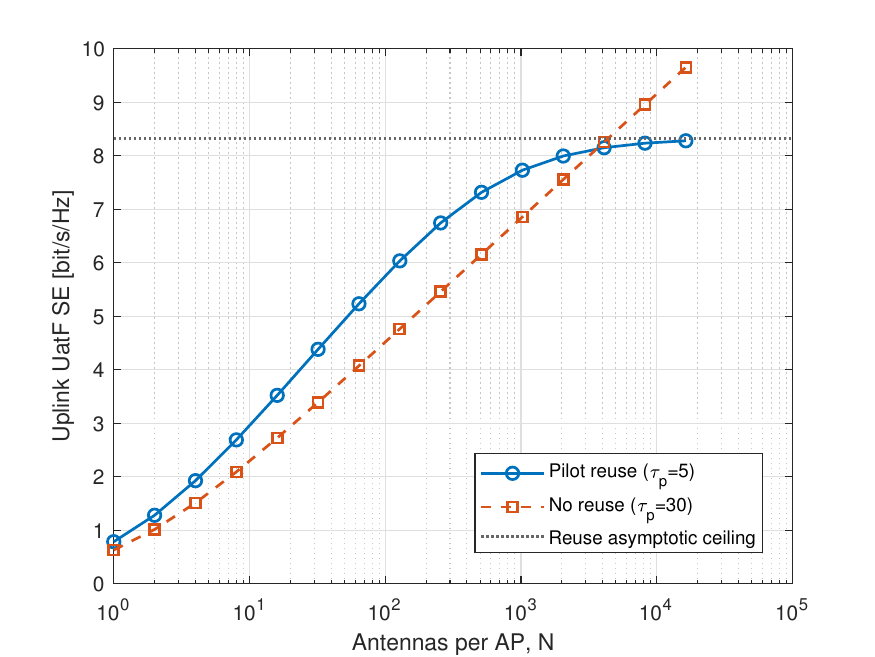}\label{fig:fig2b}}%
\caption{Validation and large-$N$ characteristics of the coordinated Q-JCKF UatF analysis.}
\label{fig:fig2}
\end{figure}

\looseness=-1
Fig.~\ref{fig:fig2} evaluates the coordinated Q-JCKF UatF analysis. In Fig.~\ref{fig:fig2a}, $K=K_r=6$, $\tau_p=1$, $L=4$, $\lambda_k=5/6$, and 2-bit ADCs are used, with $N\in\{1,2,4,8,16,32\}$. Direct conditional Monte Carlo agrees with the closed-form SINR and SE expressions in \eqref{eq:sinr_c}--\eqref{eq:qjc_uatf_rate}, with maximum relative discrepancies of approximately $1.25\%$ and $0.604\%$, respectively. Fig.~\ref{fig:fig2b} then tests the predicted large-$N$ behavior for $K=30$. With pilot reuse, $(\tau_p,K_r,\eta_u)=(5,6,0.95)$, the desired coherent term and coherent same-pilot interference have the same dominant $N$-scaling, producing the predicted finite SINR ceiling. Without reuse, $(\tau_p,K_r,\eta_u)=(30,1,0.70)$, the coherent same-pilot term vanishes, and the SINR grows approximately linearly with $N$ over the simulated range. Thus, removing coherent pilot contamination avoids the asymptotic ceiling at the cost of reducing the payload prelog from $0.95$ to $0.70$.

\begin{figure*}[!t]
\centering
\subfloat[Per-user UatF SE CDF at $2$-bit ADC resolution.]{\includegraphics[width=0.31\textwidth]{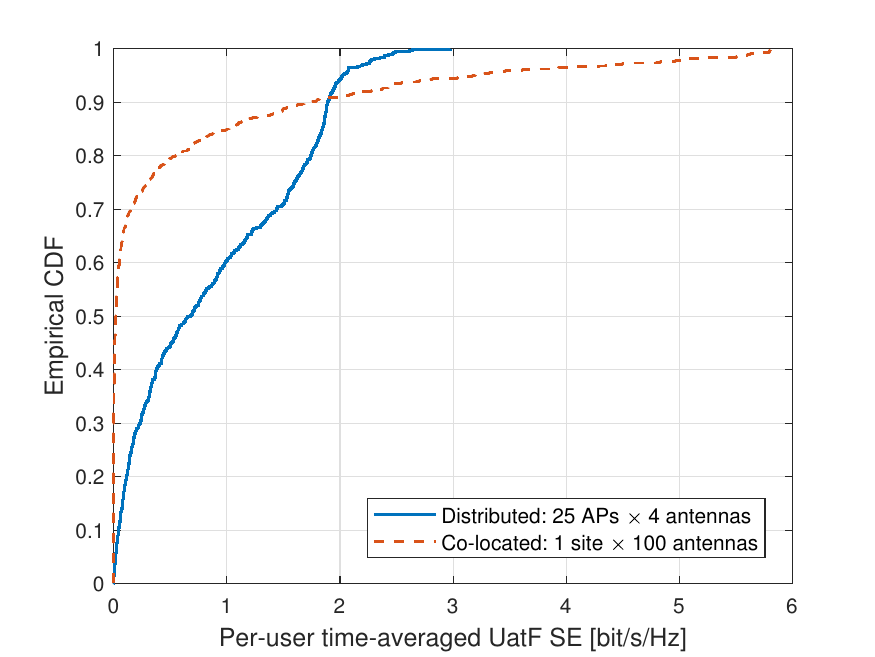}\label{fig:fig3a}}%
\hfill
\subfloat[$5$th-percentile UatF SE versus ADC resolution.]{\includegraphics[width=0.31\textwidth]{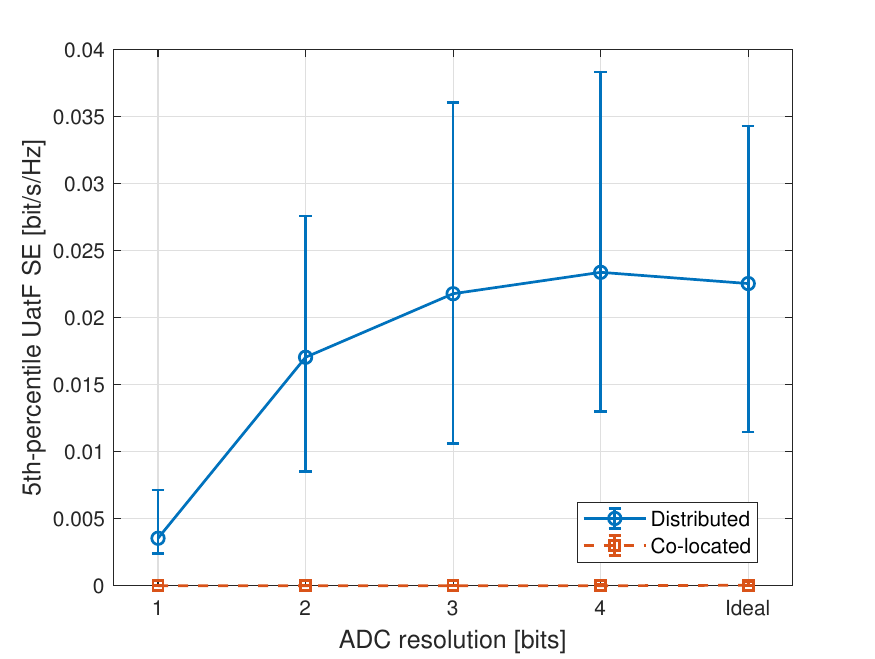}\label{fig:fig3b}}%
\hfill
\subfloat[Percentage of ideal-ADC mean UatF SE retained.]{\includegraphics[width=0.31\textwidth]{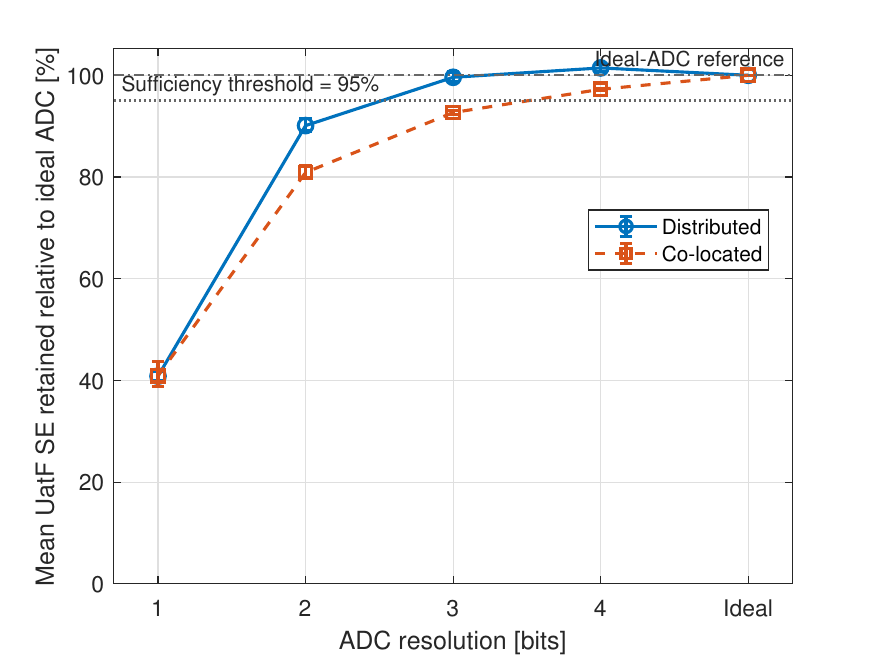}\label{fig:fig3c}}%
\caption{Distributed versus co-located deployment and ADC-resolution effects under equal total antenna count.}
\label{fig:fig3}
\end{figure*}

\looseness=-1
Fig.~\ref{fig:fig3} quantifies the effects of antenna distribution and ADC resolution for Q-JCKF under equal total antenna count. The distributed configuration uses $(L,N)=(25,4)$, whereas the centered co-located benchmark uses $(1,100)$. Results are obtained from 30 independent layouts with 8-dB log-normal shadowing and paired user locations and pilot-activity realizations. Fig.~\ref{fig:fig3a} shows that, with 2-bit ADCs, antenna distribution mainly improves the lower tail of the per-user time-averaged UatF SE through AP--user proximity and macro-diversity, whereas the centered array yields a wider rate distribution and can favor users near the array. Fig.~\ref{fig:fig3b} characterizes the corresponding 5th-percentile SE over 1--4-bit and ideal ADCs. The $95\%$ intervals are obtained from 1000 percentile cluster-bootstrap replicates, with network layouts resampled as clusters. At 2 bits, the distributed 5th-percentile SE is approximately $0.0171$ bit/s/Hz and its $95\%$ interval excludes zero, whereas the co-located lower tail is essentially zero for the centered geometry. For the distributed architecture, the 4-bit and ideal-ADC results are not practically distinguishable within the observed bootstrap uncertainty. Fig.~\ref{fig:fig3c} shows the same trend in mean SE: the fraction of ideal-ADC SE retained exhibits diminishing returns with increasing resolution. The $95\%$ reference is used only as an operating-point-dependent engineering criterion; its point estimate and lower confidence limit are reported to avoid treating it as a universal bit-resolution threshold.

\begin{figure}[!t]
\centering
\subfloat[Tracking evolution.]{\includegraphics[width=0.47\columnwidth]{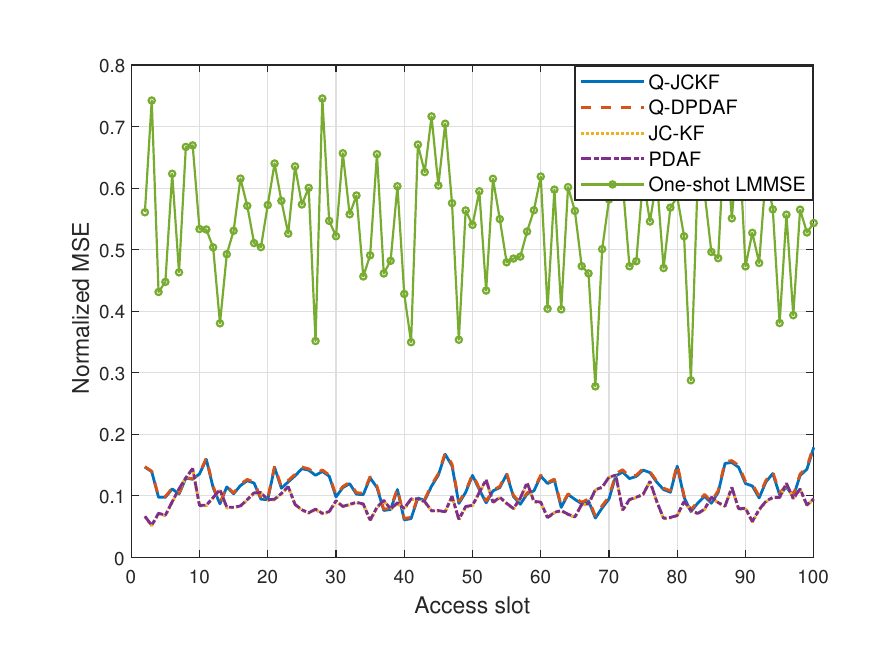}\label{fig:fig4a}}%
\hfill
\subfloat[Mobility sensitivity.]{\includegraphics[width=0.47\columnwidth]{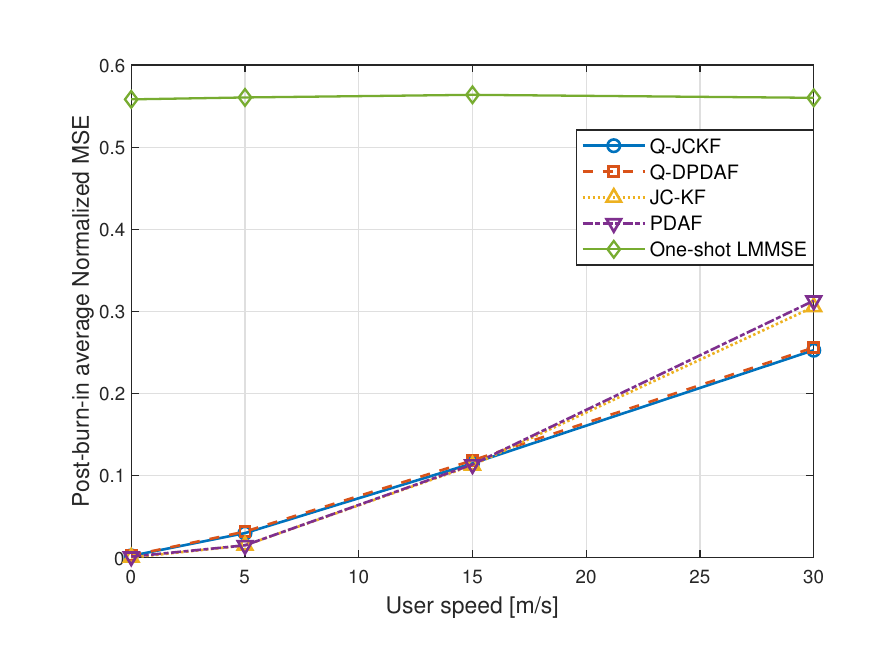}\label{fig:fig4b}}%
\caption{Channel-tracking under user mobility for the proposed distributed Q-JCKF and Q-DPDAF, the equal-antenna co-located JC-KF and PDAF baselines, and the distributed oracle-current-activity one-shot LMMSE estimator.}
\label{fig:fig4}
\end{figure}

\looseness=-1
Fig.~\ref{fig:fig4} assesses mobility robustness by comparing distributed Q-JCKF/Q-DPDAF with co-located JC-KF/PDAF and an oracle-current-activity one-shot LMMSE reference. The distributed schemes use $(L,N)=(25,4)$ with 2-bit ADCs, whereas the legacy schemes use $M=100$ ideal-ADC antennas and are applied independently to each $K_r=6$ pilot-sharing group. User locations, pilot grouping, activity statistics, transmit powers, and temporal channel models are matched across architectures, while large-scale fading is fixed within each realization. The one-shot estimator uses the same distributed 2-bit observations and the true current pilot activity, but no temporal prediction or state carryover, thus isolating the contribution of recursive tracking. For Fig.~\ref{fig:fig4a}, instantaneous network NMSE is averaged over 10 independent 100-slot realizations at $v=15$ m/s, with the common initialization slot omitted. For Fig.~\ref{fig:fig4b}, five paired 120-slot realizations are generated at each speed; the first 30 slots are discarded, and NMSE is averaged over the remaining 90 slots and then across realizations. Pairing preserves the geometry, pilot grouping, activity realization, and random-stream policy for each Monte Carlo index across the speed sweep. All recursive schemes remain well below the one-shot estimator in Fig.~\ref{fig:fig4a}. Fig.~\ref{fig:fig4b} further reveals a mobility-dependent architectural crossover. At 0 m/s, Q-JCKF/Q-DPDAF attain NMSEs of approximately $0.0020/0.0025$, while JC-KF is near zero and PDAF remains close to this level. At 5 m/s, the corresponding values are $0.0297/0.0316$ versus approximately $0.0148/0.0148$; at 15 m/s, they converge to $0.1147/0.1183$ versus $0.1123/0.1133$. At 30 m/s, the ordering reverses: Q-JCKF/Q-DPDAF achieve $0.2525/0.2558$, compared with $0.3053/0.3129$ for JC-KF/PDAF, yielding reductions of approximately $17.3\%$ and $18.2\%$ relative to the corresponding legacy baselines. Thus, the ideal-ADC co-located architecture is more accurate at low mobility, whereas the distributed quantized architecture is more robust in the considered high-mobility regime. This crossover is specific to the evaluated configurations and does not imply a universal ordering of the underlying Kalman structures. The one-shot NMSE remains approximately $0.5581$, $0.5605$, $0.5637$, and $0.5600$ at 0, 5, 15, and 30 m/s, respectively, consistent with the absence of temporal-state propagation.

\begin{figure}[!t]
\centering
\subfloat[UatF spectral efficiency.]{\includegraphics[width=0.47\columnwidth]{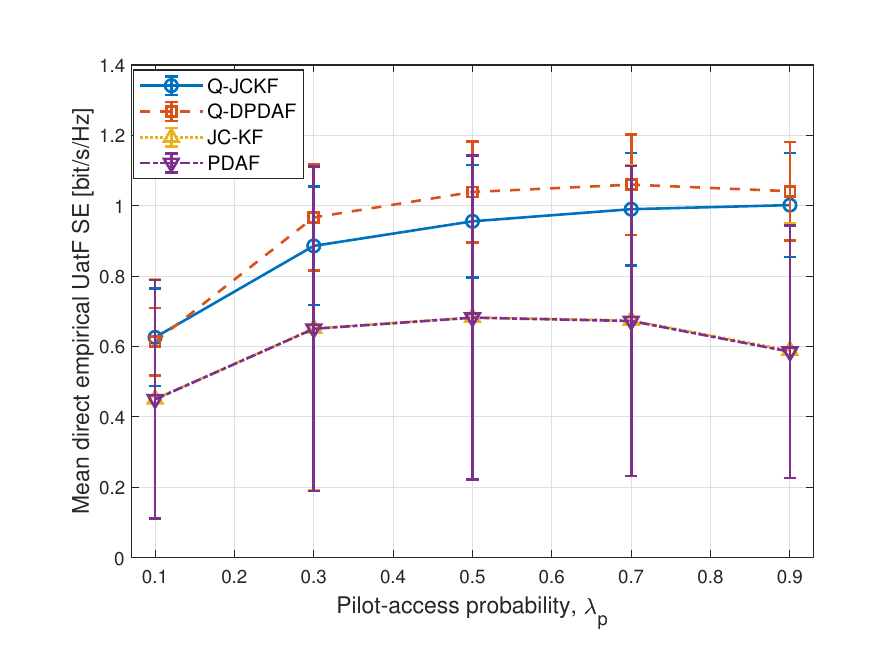}\label{fig:fig5a}}%
\hfill
\subfloat[Activity-detection accuracy.]{\includegraphics[width=0.47\columnwidth]{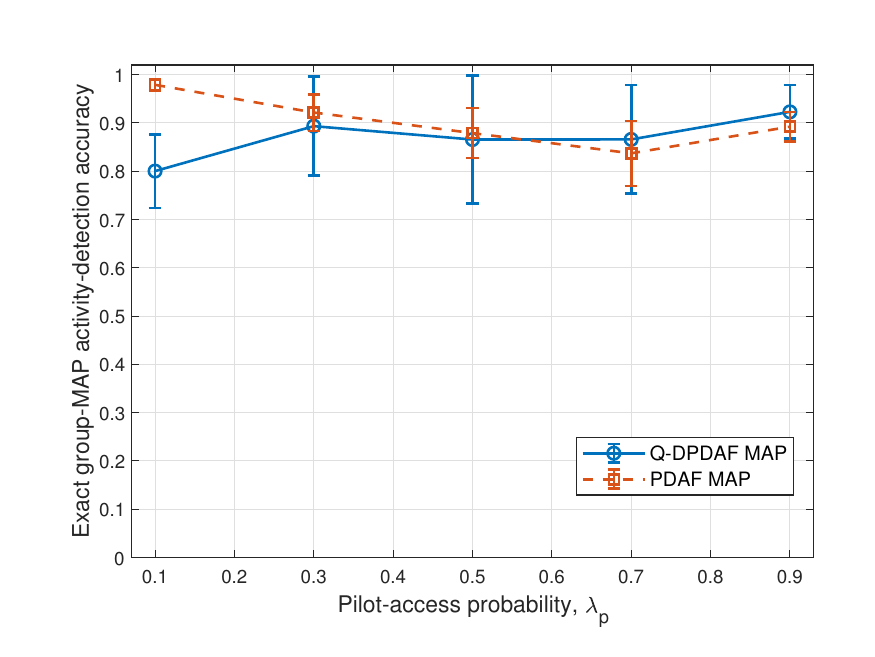}\label{fig:fig5b}}%
\caption{Effect of pilot-access probability on the proposed distributed trackers and the equal-antenna co-located legacy baselines.}
\label{fig:fig5}
\end{figure}

\looseness=-1
Fig.~\ref{fig:fig5} examines the common pilot-access probability $\lambda\in\{0.1,0.3,0.5,0.7,0.9\}$ for $K=30$, $\tau_p=5$, $K_r=6$, and $v=15$ m/s. Q-JCKF/Q-DPDAF use the distributed $(25,4)$, 2-bit architecture, whereas JC-KF/PDAF use the equal-antenna $M=100$ ideal-ADC benchmark. Shadowing is disabled, all users are data-active during rate evaluation, and results are averaged over five independent layouts with two independent replicas per layout and configuration; the error bars represent $95\%$ layout-level confidence intervals. Fig.~\ref{fig:fig5a} shows higher mean UatF SE point estimates for the distributed trackers throughout the tested $\lambda$ range. Q-JCKF increases from approximately $0.626$ bit/s/Hz at $\lambda=0.1$ to $1.002$ bit/s/Hz at $\lambda=0.9$. Q-DPDAF increases from $0.614$ bit/s/Hz to $1.060$ bit/s/Hz at $\lambda=0.7$ and then decreases slightly to $1.041$ bit/s/Hz at $\lambda=0.9$. In contrast, JC-KF and PDAF peak near $0.68$ bit/s/Hz around $\lambda=0.5$ and decrease to approximately $0.59$ bit/s/Hz at $\lambda=0.9$. This non-monotonic behavior reflects the competing effects of more frequent channel updates and increased simultaneous same-pilot activity; for the low-resolution distributed receiver, the latter also increases aggregate ADC input power and quantization distortion. The modest Q-DPDAF advantage over Q-JCKF for $\lambda\geq0.3$ should therefore not be interpreted as an intrinsic advantage of uncoordinated tracking. Fig.~\ref{fig:fig5b} reports prior-aware exact group-MAP activity-detection accuracy. Because $\lambda$ also determines the activity-pattern prior, accuracies at different $\lambda$ values are not a prior-independent detector metric. For $K_r=6$, a prior-only MAP rule would achieve accuracies of approximately $0.531$, $0.118$, $0.0156$, $0.118$, and $0.531$ over the tested $\lambda$ values. PDAF achieves $0.979$, $0.922$, $0.879$, $0.837$, and $0.892$, whereas Q-DPDAF achieves $0.800$, $0.893$, $0.866$, $0.866$, and $0.923$. Hence, PDAF gives higher point estimates at low $\lambda$, while Q-DPDAF is higher at $\lambda=0.7$ and $0.9$. This reversal reflects the combined effects of the activity prior, observation statistics, receiver architecture, quantization, and distributed likelihood information rather than AP distribution alone. The confidence intervals, particularly for the co-located SE baselines, also indicate appreciable layout sensitivity; the figure should thus be interpreted in terms of system-level trends rather than a universal statistical ordering among algorithms.

\begin{figure}[!t]
\centering
\includegraphics[width=\columnwidth]{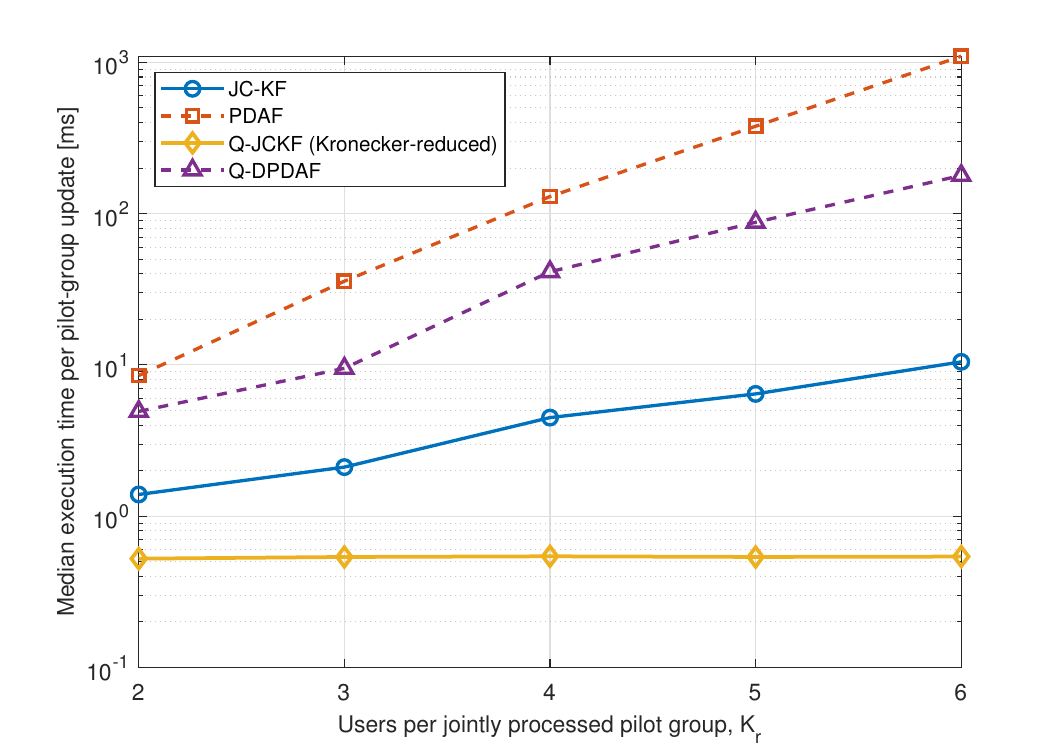}
\caption{Empirical single-process MATLAB median execution time per pilot-group update versus $K_r$ for JC-KF, PDAF, Q-JCKF, and Q-DPDAF.}
\label{fig:fig6}
\end{figure}

Fig.~\ref{fig:fig6} relates the analytical complexity results of Section~V to measured computation time. Single-process MATLAB runtimes are reported for $K_r\in\{2,3,4,5,6\}$, $\lambda_k=5/6$, and $\rho=0.95$. JC-KF/PDAF use $M=100$ ideal-ADC antennas, whereas Q-JCKF/Q-DPDAF use $(L,N)=(25,4)$ with 2-bit ADCs. Median execution times are obtained from repeated, adaptively batched measurements after JIT warm-up, excluding data generation, validation, plotting, and file I/O. AP-local computations are executed serially, so the measurements reflect single-process work rather than parallel critical-path latency.

\looseness=-1
The timed Q-JCKF exploits the exact Kronecker reduction of Section~III-A, replacing the full $NK_r\times NK_r$ covariance recursion by its $K_r\times K_r$ user-domain form. Structured and dense implementations agree over all 25 APs to relative mean and covariance discrepancies on the order of $10^{-16}$, confirming numerical equivalence to floating-point precision. Q-JCKF requires approximately $0.52$--$0.54$ ms per update over the tested group sizes; the nearly flat curve reflects fixed implementation overhead dominating the small user-domain matrix operations and should not be used to infer asymptotic scaling. PDAF and Q-DPDAF increase substantially faster because of their $2^{K_r}$ hypothesis enumeration. At $K_r=6$, the measured runtimes are $10.476$ ms for JC-KF, $1097.120$ ms for PDAF, $0.541$ ms for Q-JCKF, and $178.821$ ms for Q-DPDAF. Thus, under the reported serial implementation, Q-JCKF is approximately $19.36$ times faster than JC-KF, while Q-DPDAF is approximately $6.14$ times faster than PDAF. These ratios are implementation-dependent and provide empirical support for the complexity trends established in Section~V rather than hardware-independent speedup factors.

\looseness=-1
Taken together, Figs.~1--6 support the principal analytical and algorithmic claims of the paper. The simulations verify the coordinated and mixture-aware UatF expressions, the cross-AP association bound, and the predicted large-$N$ pilot-reuse ceiling; quantify the lower-tail benefit of distributed reception and the diminishing penalty of increasing ADC resolution; demonstrate the importance of recursive tracking under intermittent pilots and mobility; and characterize Q-DPDAF activity inference under a $\lambda$-dependent prior. The measured runtimes further show that the Kronecker reduction and pilot-group-wise formulation provide substantial computational savings. Overall, the results establish the main operating trade-off of the framework: distributed low-resolution reception can preserve favorable rate and tracking behavior while shifting the dominant processing burden from the full user population to the pilot-group dimension.

\section{Conclusion}

\looseness=-1
This paper developed quantization-aware channel tracking for coordinated and uncoordinated access in distributed mMIMO with low-resolution ADCs. Q-JCKF exploits known pilot activity through an exact Kronecker-structured covariance reduction, whereas Q-DPDAF for uncoordinated access combines AP-local hypothesis-conditioned Kalman updates with CPU-based probabilistic activity fusion and Gaussian moment matching. For uplink performance evaluation, we derived a closed-form statistical UatF expression for Q-JCKF, together with mixture-aware, pilot-conditioned, and statistical UatF formulations for Q-DPDAF. The analysis established a finite large-array SINR ceiling under persistent coherent pilot-reuse interference and characterized the current-slot cross-AP association induced by shared activity uncertainty, including its $\mathcal{O}(\epsilon)$ posterior-concentration bound. We also recovered the ideal-ADC co-located JC-KF/PDAF model as a limiting case and quantified the conditional loss of Kalman measurement information due to finite-resolution quantization. From a scalability perspective, the Q-JCKF Kronecker structure reduces covariance recursion to the $K_{r}$-dimensional user domain, while groupwise Q-DPDAF limits hypothesis enumeration from $2^{K}$ to $2^{K_{r}}$ and supports parallel AP-local processing with lower processing and fronthaul requirements. Numerical results corroborated the UatF analysis, large-N behavior, association bound, and complexity trends, while showing improved lower-tail spectral efficiency, near-ideal performance at moderate ADC resolution, and robust tracking under mobility and intermittent pilot access.

\end{document}